\documentclass[a4paper,fleqn]{cas-dc}
\usepackage{amsmath,lipsum}
\usepackage{cuted}%%stripsep-3pt
\usepackage{lineno}
\usepackage{booktabs,makecell,multirow}
\usepackage{amsmath,amssymb,amsfonts}
\usepackage{algorithmic}
\usepackage{graphicx}
\usepackage{textcomp}
\usepackage{amsmath,amsfonts}
\usepackage{algorithmic}
\usepackage{algorithm}
\usepackage{array}
\usepackage[caption=false,font=normalsize,labelfont=sf,textfont=sf]{subfig}
\usepackage{textcomp}
\usepackage{stfloats}
\usepackage{url}
\usepackage{verbatim}
\usepackage{graphicx}
\usepackage{cite}
\usepackage{booktabs}

\usepackage{upgreek}

\usepackage{threeparttable}

\usepackage[numbers,sort&compress]{natbib}

\def\tsc#1{\csdef{#1}{\textsc{\lowercase{#1}}\xspace}}
\tsc{WGM}
\tsc{QE}
\usepackage{xspace}
\makeatletter
\DeclareRobustCommand\onedot{\futurelet\@let@token\@onedot}
\def\@onedot{\ifx\@let@token.\else.\null\fi\xspace}

\makeatother

\usepackage{color}

\usepackage{caption}
\begin{document}
\let\WriteBookmarks\relax
\def\floatpagepagefraction{1}
\def\textpagefraction{.001}

\shorttitle{Design, analysis, and manufacturing of a glass-plastic hybrid minimalist aspheric panoramic annular lens}    

\shortauthors{S. Gao \textit{et al.}}  

\title [mode = title]{Design, analysis, and manufacturing of a glass-plastic hybrid minimalist aspheric panoramic annular lens}

\author[1,4]{Shaohua Gao}
\credit{Conceptualization of this study, Methodology, Software}
\author[1,4]{Qi Jiang}
\author[1]{Yiqi Liao}
\author[2]{Yi Qiu}
\author[2]{Wanglei Ying}
\author[3]{Kailun Yang}
\author[1,4]{Kaiwei Wang}[orcid=0000-0002-8272-3119]
\cormark[1]
\ead{wangkaiwei@zju.edu.cn}
\author[5]{Benhao Zhang}
\author[1]{Jian Bai}

\affiliation[a]{organization={State Key Laboratory of Extreme Photonics and Instrumentation, College of Optical Science and Engineering},
            addressline={Zhejiang University}, 
            city={Hangzhou},
            postcode={310027}, 
            country={China}}

\affiliation[b]{organization={Ningbo Lian Technology Co., Ltd},
            addressline={Ningbo Lian},
            city={Ningbo},
%          citysep={}, % Uncomment if no comma needed between city and postcode
            postcode={315500}, 
            country={China}}

\affiliation[c]{organization={National Engineering Research Center of Robot Visual Perception and Control Technology},
            addressline={Hunan University}, 
            city={Changsha},
%          citysep={}, % Uncomment if no comma needed between city and postcode
            postcode={410082}, 
            country={China}}

\affiliation[c]{organization={Intelligent Optics $\&$ Photonics Research Center, Jiaxing Research Institute},
            addressline={Zhejiang University}, 
            city={Jiaxing},
%          citysep={}, % Uncomment if no comma needed between city and postcode
            postcode={314031}, 
            country={China}}

\affiliation[c]{organization={Central Research Institue of Sunny Optical Technology},
            addressline={Sunny Optical Technology}, 
            city={Hangzhou},
%          citysep={}, % Uncomment if no comma needed between city and postcode
            postcode={311215}, 
            country={China}}

\cortext[1]{Corresponding author}



\begin{abstract}
We propose a high-performance glass-plastic hybrid minimalist aspheric panoramic annular lens (ASPAL) to solve several major limitations of the traditional panoramic annular lens (PAL), such as large size, high weight, and complex system. The field of view (FoV) of the ASPAL is $360^\circ{\times}(35^\circ{\sim}110^\circ$) and the imaging quality is close to the diffraction limit. This large FoV ASPAL is composed of only 4 lenses. Moreover, we establish a physical structure model of PAL using the ray tracing method and study the influence of its physical parameters on compactness ratio. In addition, for the evaluation of local tolerances of annular surfaces, we propose a tolerance analysis method suitable for ASPAL. This analytical method can effectively analyze surface irregularities on annular surfaces and provide clear guidance on manufacturing tolerances for ASPAL. Benefiting from high-precision glass molding and injection molding aspheric lens manufacturing techniques, we finally manufactured 20 ASPALs in small batches. The weight of an ASPAL prototype is only 8.5 g. Our framework provides promising insights for the application of panoramic systems in space and weight-constrained environmental sensing scenarios such as intelligent security, micro-UAVs, and micro-robots.
\end{abstract}

% Keywords
\begin{keywords}
\sep Lens system design
 \sep Optical design of instruments
 \sep Optical instruments and components
 \sep Panoramic imaging system
 \sep Panoramic annular lens
\end{keywords}
\maketitle

% Main text
\section{Introduction}\label{1}

With the prosperous development of precision optical manufacturing technology and optical sensing field, optical systems are widely used as an environment sensing device and are expected to have the advantages of large field of view (FoV), high performance, and miniaturization~\cite{miller2023optics,zhang2023large}. The panoramic imaging system is a special optical system that can acquire a large FoV with a maximum FoV of more than 180$^{\circ}$ of the surrounding environment. With this unique imaging characteristic, it has broader application prospects in video meetings, intelligent security, environment perception, panoramic photography, virtual reality/augmented reality panoramic scene acquisition, machine vision, endoscopic detection, miniaturized robot perception, drone perception, and other related fields~\cite{gao2022review}.
To realize the perceptual ability of panoramic imaging systems, several practical designs, such as fisheye lens, catadioptric panoramic system, hyper-hemispheric panoramic lens, and panoramic annular lens (PAL) have been gradually proposed, developed, and applied~\cite{pernechele2021telecentric, Arkhipova2016CircularscanPS, Pernechele2013HyperhemisphericAB, wang2022high}, for obtaining large FoV and high-quality imaging in the real world.
In terms of large FoV panoramic imaging, catadioptric panoramic architectures, and entrance pupil preposition architectures have been successively proposed~\cite{Yang:23, wang2022design, Pan:23,ke2023ultra,wang2015design}. Cheng~\textit{et al.} designed and fabricated an ultrawide angle catadioptric lens by using an annular stitched aspheric surface type~\cite{Cheng2016DesignOA}. The maximum FoV of the panoramic system reaches $360^\circ{\times}135^\circ$. 
Zhang~\textit{et al.} proposed a dual-mirror catadioptric panoramic architecture to obtain a large $360^\circ{\times}(22^\circ{\sim}120^\circ$) FoV and a small blind area~\cite{Zhang2020DesignOA}. 
Wang~\textit{et al.} proposed an entrance pupil preposition panoramic annular lens with an ogive surface to achieve a long focal length panoramic system~\cite{wang2015design}. 
Huang~\textit{et al.} developed a dual-channel FoV entrance pupil preposition panoramic annular lens in the front and rear directions~\cite{huang2017design}. In terms of miniaturization, Q-type surfaces, and freeform surfaces have been successively used in the panoramic lens design to improve the compactness of the panoramic system~\cite{Gao2021DesignOA, Zhu:22,gao2022compact}. 
In terms of achieving high performance, panoramic lens designs with spherical and aspherical surfaces have been proposed to realize high quality designs close to the diffraction limit and low distortion designs, respectively~\cite{wang2019design, Zhou2020DesignAI}.
The above studies have made outstanding research for realizing panoramic imaging systems in terms of large FoV, miniaturization, and high performance. However, the current proposed panoramic systems still cannot simultaneously address the drawbacks of large volume, heavy weight, and complex systems. The complexity of the panoramic systems often results in a complex assembly process, which in turn leads to difficulties in the manufacture and assembly of panoramic systems. It is not conducive to volume industrial manufacturing and applications.
In addition, it is not possible to reasonably assess the sag deviation of aspheric surfaces in the manufacturing process according to the standard surface tolerance analysis method. 
In this way, it will result in the difficulty of reasonably providing the manufacturer with the surface accuracy requirements, which ultimately leads to poor image quality or overly tight tolerance requirements with high costs.
Moreover, evaluating the local tolerances of optical surfaces is particularly important for manufacturing high-precision optical imaging systems~\cite{deng2022local}. Due to the annular optical surfaces of the PALs, it is necessary to evaluate the figure errors of the annular surfaces, rather than the full aperture figure errors.

To solve the problem that traditional PALs suffer large sizes, high weight, and complex systems in pursuit of high-performance imaging quality and large FoV panoramic imaging, we propose a high-performance glass-plastic hybrid minimalist aspheric panoramic annular lens (ASPAL). The ASPAL consists of only 4 lenses. It weighs only 8.5 g and is smaller than a coin. For convincing evaluation of the local tolerance of the surfaces, we propose an aspheric surface tolerance analysis method for the ASPAL. The method can reasonably evaluate the figure errors of the annular surfaces and the non-annular surfaces of the ASPAL, including the peak-to-valley (PV) and root-mean-square (RMS) figure errors. Finally, we fabricated 20 ASPALs and verified their imaging performance. The development principles of the ASPAL are described in Section~\ref{sec:2}.

\section{Development and ray tracing compact PAL architecture}
\label{sec:2}
\subsection{Comparison of different panoramic imaging system architecture designs and compactness}

To design a compact panoramic imaging system, it is necessary to analyze the compactness of various panoramic imaging systems. To rationally characterize the compactness ratio of a panoramic imaging system, the three compactness categories are defined as Eq.~(\ref{eq:1}) :

\begin{equation}
\label{eq:1}
\left\{\begin{array}{l}
CR_1=\frac{D_{\max }}{TL} \\
CR_2=\frac{D_{\max }}{D_{\min }} \\
CR_2=\frac{D_{\max} }{D_{IMA}}
\end{array}\right.
{.}
\end{equation}

$CR_1$ is the ratio of the maximum lens diameter $D_{\max}$ to the total length $TL$. $CR_2$ is the ratio of the maximum lens diameter $D_{\max}$ to the minimum lens diameter $D_{\min}$. $CR_3$ is the ratio of the maximum lens diameter $D_{\max}$ to the maximum imaging circle diameter $D_{IMA}$.
We conducted comparative designs and compactness analysis of seven different panoramic imaging architectures including a fisheye lens, a catadioptric system with a single mirror, a catadioptric system with two mirrors in the same direction, a catadioptric system with two mirrors in opposite directions, a hyper-hemispheric lens, an entrance pupil preposition-panoramic annular lens, and a panoramic annular lens. Different types of optical systems have different advantages and disadvantages. Optical design typically involves balancing various parameters when designing panoramic imaging systems.

\begin{figure*}[ht!]
\centering\includegraphics[width=13.0cm]{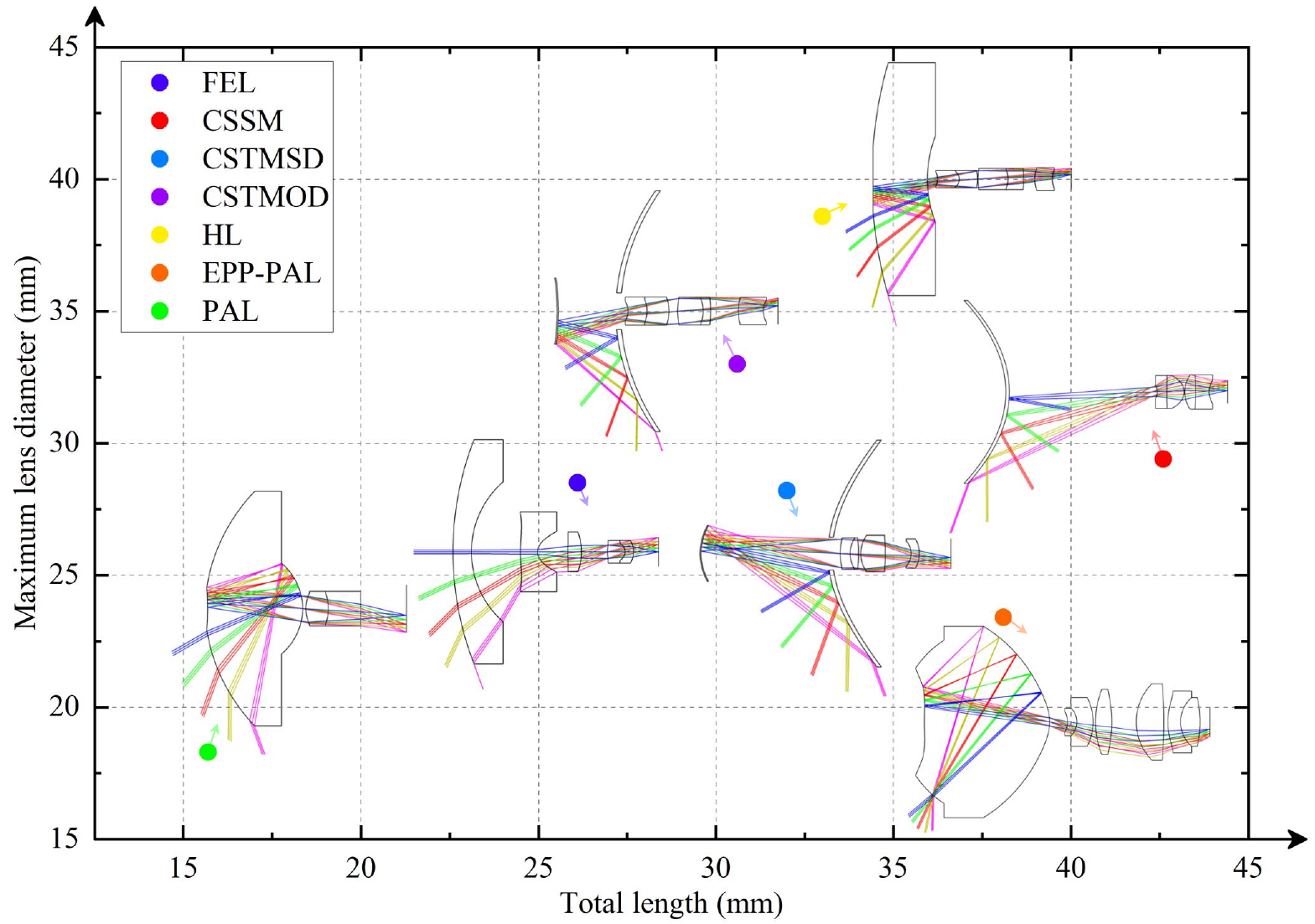}
\caption{The optical path diagrams and coordinate distribution of maximum lens diameter $D_{\max}$ and total length $TL$ of seven panoramic imaging systems designed with an imaging circle diameter of 3.6 mm as the design criteria. Compared with other imaging systems, PAL has significant advantages in maximum lens diameter and total length and has the least number of lenses.}
\label{fig1}
\vskip-5ex
\end{figure*}

To compare the compactness and wide FoV perception advantages of different panoramic imaging systems reasonably and fairly, the design criteria for these panoramic systems include visible light, the same 3.6 mm imaging circle diameter, the closest possible FoV, the smallest possible $F$$/$\#, the best imaging quality, and the least possible number of lenses. The final imaging performance evaluation indicators include spot radius, modulation transfer function, distortion, relative illumination, and chief ray angle. For the calculation of distortion, except for the imaging relationship of the entrance pupil preposition-panoramic annular lens is $y = f \theta-h$, all other panoramic imaging systems meet the equidistance project model of $y = f\theta$. Here, $y$ is the image height, $f$ is the focal length, $\theta$ is the angle of incidence, and $h$ is the compression value~\cite{wang2015design}. The final designed system indicators are shown in Table~\ref{tab:1}. Under the same imaging circle diameter of 3.6 mm and imaging quality as close as possible, PAL has a clear advantage of a smaller maximum lens diameter and total length, as shown in Fig.~\ref{fig1}. In addition, PAL has significantly smaller compactness ratios of 2 and 3, indicating that it has a more compact structure. Considering the factor of low-cost mass production, PAL can achieve imaging perception of compact and ultra-large FoV using only 3 plastic lenses, which has the potential for low-cost mass production and application prospects. However, the advantage of PAL's compactness ratio $CR_1$ is not obvious, so it is necessary to analyze the physical parameters of PAL that affect the structural compactness to achieve a more compact structural compactness ratio $CR_1$.

\begin{table*}[!ht]
\centering
\scriptsize
\begin{threeparttable}
\caption{Comparison of parameters for different panoramic imaging system architectures.\tnote{\textit{a}}}
\setlength{\tabcolsep}{0.5mm}{
\begin{tabular}{cccccccccccccccc}
\toprule
\makecell[c]{System
}&\makecell[c]{FL
(mm)
}& \makecell[c] {HFoV
($^{\circ}$)}&\makecell[c]{F/\#
}&\makecell[c]{SR ($\upmu$m)}&\makecell[c]{MTF (lp/mm)}&\makecell[c]{D ($\%$)}&\makecell[c]{RI}&\makecell[c]{CRA ($^{\circ}$)}&\makecell[c]{MLD (mm)}&\makecell[c]{TL (mm)}&\makecell[c]{CR$_1$}&\makecell[c]{CR$_2$}&\makecell[c]{CR$_3$}&\makecell[c]{SA}\\
\midrule
\makecell[c]{FEL}&\makecell[c]{1.14
}&\makecell[c]{0$^\circ$ $\sim $110$^\circ$}&\makecell[c]{2.5}&\makecell[c]{2.7}&\makecell[c]{0.42@133lp/mm
}&\makecell[c]{-15.0}&\makecell[c]{0.75}&\makecell[c]{16}&\makecell[c]{28.5}&\makecell[c]{26.1}&\makecell[c]{1.09}&\makecell[c]{12.7}&\makecell[c]{7.9}&\makecell[c]{2G3P}\\
\midrule
\makecell[c]{CSSM}&\makecell[c]{1.09
}&\makecell[c]{10$^\circ$ $\sim $110$^\circ$}&\makecell[c]{2.8}&\makecell[c]{4.6}&\makecell[c]{0.56@133lp/mm
}&\makecell[c]{-14.0}&\makecell[c]{0.87}&\makecell[c]{9}&\makecell[c]{29.4}&\makecell[c]{42.6}&\makecell[c]{0.69}&\makecell[c]{6.9}&\makecell[c]{8.2}&\makecell[c]{1M4G}\\
\midrule
\makecell[c]{CSTMSD}&\makecell[c]{-1.00
}&\makecell[c]{30$^\circ$ $\sim $110$^\circ$}&\makecell[c]{3}&\makecell[c]{2.8}&\makecell[c]{0.56@133lp/mm
}&\makecell[c]{0.3}&\makecell[c]{0.95}&\makecell[c]{10}&\makecell[c]{28.2}&\makecell[c]{32.0}&\makecell[c]{0.88}&\makecell[c]{7.6}&\makecell[c]{7.8}&\makecell[c]{2M4G}\\
 \midrule
\makecell[c]{CSTMOD}&\makecell[c]{1.50
}&\makecell[c]{30$^\circ$ $\sim $110$^\circ$}&\makecell[c]{3}&\makecell[c]{2.3}&\makecell[c]{0.47@133lp/mm
}&\makecell[c]{-37.3}&\makecell[c]{0.88}&\makecell[c]{22}&\makecell[c]{33.0}&\makecell[c]{30.6}&\makecell[c]{1.08}&\makecell[c]{10.2}&\makecell[c]{9.2}&\makecell[c]{2M5G}\\
\midrule
\makecell[c]{HL}&\makecell[c]{1.47
}&\makecell[c]{30$^\circ$ $\sim $105$^\circ$}&\makecell[c]{3.5}&\makecell[c]{2.6}&\makecell[c]{0.41@133lp/mm
}&\makecell[c]{-34.1}&\makecell[c]{0.97}&\makecell[c]{12}&\makecell[c]{38.6}&\makecell[c]{33.0}&\makecell[c]{1.17}&\makecell[c]{14.4}&\makecell[c]{10.7}&\makecell[c]{6G}\\
\midrule
\makecell[c]{EPP-PAL}&\makecell[c]{-1.78
}&\makecell[c]{40$^\circ$ $\sim $89$^\circ$}&\makecell[c]{5.5}&\makecell[c]{3.8}&\makecell[c]{0.35@100lp/mm
}&\makecell[c]{-3.3}&\makecell[c]{0.66}&\makecell[c]{27}&\makecell[c]{23.4}&\makecell[c]{38.1}&\makecell[c]{0.61}&\makecell[c]{6.9}&\makecell[c]{6.5}&\makecell[c]{7G}\\
\midrule
\makecell[c]{PAL}&\makecell[c]{-1.01
}&\makecell[c]{30$^\circ$ $\sim $110$^\circ$}&\makecell[c]{3.1}&\makecell[c]{2.6}&\makecell[c]{0.55@133lp/mm
}&\makecell[c]{-4.6}&\makecell[c]{0.80}&\makecell[c]{16}&\makecell[c]{\textbf{18.3}}&\makecell[c]{\textbf{15.7}}&\makecell[c]{1.17}&\makecell[c]{\textbf{6.8}}&\makecell[c]{\textbf{5.1}}&\makecell[c]{\textbf{3P}}\\

\bottomrule
\label{tab:1}
\end{tabular}}

\begin{tablenotes}
        \footnotesize
        \item[\textit{a}] FEL: Fisheye Lens. CSSM: Catadioptric System with a Single Mirror. CSTMSD: Catadioptric System with Two Mirrors in the Same Direction. CSTMOD: Catadioptric System with Two Mirrors in Opposite Directions. HL: Hyper-hemispheric lens. EPP-PAL: Entrance Pupil Preposition-Panoramic Annular Lens. PAL: Panoramic Annular Lens. FL: Focal Length. HFoV: Half Field of View. F/\#: F-number. SR: Spot Radius. MTF: Modulation Transfer Function. D: Distortion. RI: Relative Illumination. CRA: Chief Ray Angle. MLD: Maximum Lens Diameter. TL: Total Length. CR$_1$: Compactness Ratio 1. CR$_2$: Compactness Ratio 2. CR$_3$: Compactness Ratio 3. SA: System Architecture. M: Mirror. G: Glass. P: Plastic.
    
      \end{tablenotes}
      \small
    \end{threeparttable}
\end{table*}

\subsection{Ray tracing analysis of PAL physical parameters for structural compactness}
\label{sec:2.2}
Traditional optical design software can calculate and optimize various optical systems. However, these optical design software tools are unable to study the trend of changes in physical structural parameters that affect the compactness of optical systems, which makes it necessary for us to construct ray tracing models in numerical analysis software and accurately simulate the physical structure of PAL. We introduce the ray tracing equation based on Snell's law to construct a ray tracing model for PAL. According to Snell's Law, the Refractive Law~\cite{li2014geometrical} can be expressed as Eq.~(\ref{eq:2}).

\begin{equation}
\label{eq:2}
n_2 \sin \theta_2=n_1 \sin \theta_1
{.}
\end{equation}

In Eq.~(\ref{eq:2}), $n_1$ and $n_2$ are the refractive indices of medium 1 and medium 2, respectively. $\theta_1$ and $\theta_2$ are the incident angle and refractive angle, respectively. The unit vector of the interface normal between medium 1 and medium 2 is taken as $\vec{\Omega}$, with the direction perpendicular to the interface direction, pointing from medium 1 to medium 2. The unit vectors in two media along the direction of the ray are set to be $\vec{p_1}$ and $\vec{p_2}$. Eq.~(\ref{eq:2}) can be written as the vector form in Eq.~(\ref{eq:3}):

\begin{equation}
\label{eq:3}
n_2 \vec{p_2} \times \vec{\Omega}=n_1 \vec{p_1} \times \vec{\Omega}
{.}
\end{equation}

Introducing $\vec{k}=n \vec{p}$, Eq.~(\ref{eq:3}) can be expressed as Eq.~(\ref{eq:4}):

\begin{equation}
\label{eq:4}
\left(\vec{k_2}-\vec{k_1}\right) \times \vec{\Omega}=\vec{0}
{.}
\end{equation}

It can be inferred that $\vec{k_2}-\vec{k_1}$ and $\vec{\Omega}$ are parallel. Therefore, a proportional coefficient $\xi$ can be set in Eq.~(\ref{eq:5}).

\begin{equation}
\label{eq:5}
\vec{k_2}=\vec{k_1}+\xi \vec{\Omega}
{.}
\end{equation}

By using $\vec{\Omega}$ to dot multiply both sides of Eq.~(\ref{eq:5}) simultaneously. Eq.~(\ref{eq:6}) can be obtained.

\begin{equation}
\label{eq:6}
\xi=\vec{k_2} \cdot \vec{\Omega}-\vec{k_1} \cdot \vec{\Omega}
{.}
\end{equation}

Because $\left(\vec{k_2} \cdot \vec{\Omega}\right)^2=n_2^2-n_1^2+\left(\vec{k_1} \cdot \vec{\Omega}\right)^2$, for the refractive state of ray tracing, there is Eq.~(\ref{eq:7}):

\begin{equation}
\label{eq:7}
\left\{\begin{array}{l}
\vec{k_2} \cdot \vec{\Omega}=\sqrt{n_2^2-n_1^2+\left(\vec{k_1} \cdot \vec{\Omega}\right)^2} \\
\xi=\sqrt{n_2^2-n_1^2+\left(\vec{k_1} \cdot \vec{\Omega}\right)^2}-\vec{k_1} \cdot \vec{\Omega}
\end{array}\right.
{.}
\end{equation}

If the ray is in a reflective state, $n_2 = -n_1$, $\theta_2 =-\theta_1$. Then, there is Eq.~(\ref{eq:8}):

\begin{equation}
\label{eq:8}
\left\{\begin{array}{l}
\vec{k_2} \cdot \vec{\Omega}=-\vec{k_1} \cdot \vec{\Omega} \\
\xi=-2 \vec{k_1} \cdot \vec{\Omega}
\end{array}\right.
{.}
\end{equation}

\begin{figure*}[ht!]
\centering\includegraphics[width=14cm]{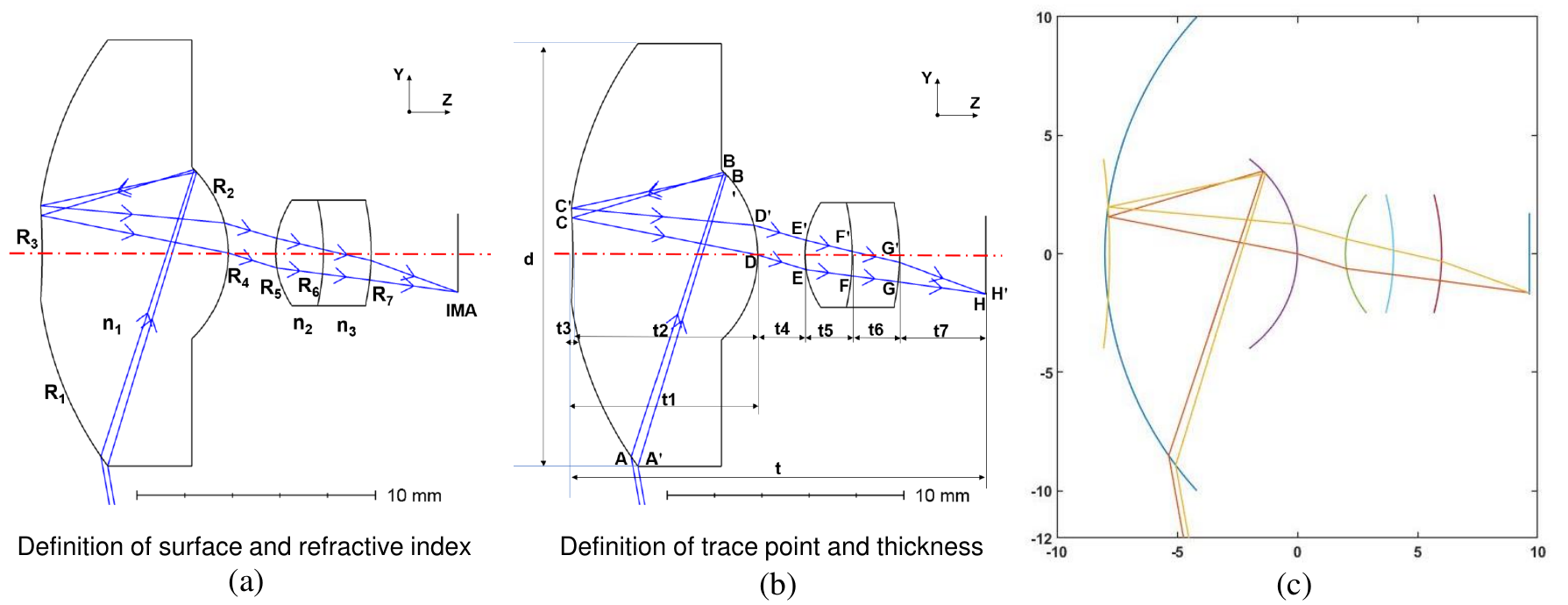}
\caption{Establishment of PAL physical structure model and implementation of ray tracing. (a) The surface name and refractive index definition of the imaging rays passing through PAL. (b) The tracking points and surface thickness definition of the marginal and chief rays passing through the interior of PAL. (c) Physical structure model of PAL in Matlab constructed by point-by-point ray tracing.}
\label{fig2}
\vskip-5ex
\end{figure*}

Combining Eq.~(\ref{eq:5}), Eq.~(\ref{eq:7}), and Eq.~(\ref{eq:8}) can calculate the direction of refraction ray or reflection ray based on refractive index, the direction of the incident light, and normal direction, and surface curvature. To perform parameter modeling and ray tracing analysis on the physical structure of PAL, we analyze its compactness. The compactness of PAL is usually related to the coordinate axis positions of the first incident point and the focal point on the final image plane. For calculating the physical structural parameters of PAL and the positions of each point, we perform ray tracing on its main and marginal rays in the Y-Z Cartesian coordinate system. 
Aspheric surfaces bring more optimization degrees of freedom to correct primary and higher-order aberrations. The fact that the coefficients of the aspheric surface are not orthogonal causes the aspheric parameters to interact with each other. Therefore, we construct the initial structure of PAL using the Petzval sum correction method with standard spherical surfaces. Firstly, define the surface and refractive index through which the incident light passes, as shown in Fig.~\ref{fig2}(a). Secondly, we define the thickness between the points and the surface that the incident light passes through sequentially, as shown in  Fig.~\ref{fig2}(b). By defining the initial structure and iteratively calculating the ray tracing of the PAL optical path, Fig.~\ref{fig2}(c) was obtained. The visualization effect of point-by-point tracing of the ray path in Matlab and the precise coordinate values of each point are completely consistent with the coordinate point positions in the optical design software. By tracking the coordinate values point by point, we can also calculate the physical structure parameters of the PAL compactness ratio.

\begin{figure*}[ht!]
\centering\includegraphics[width=16cm]{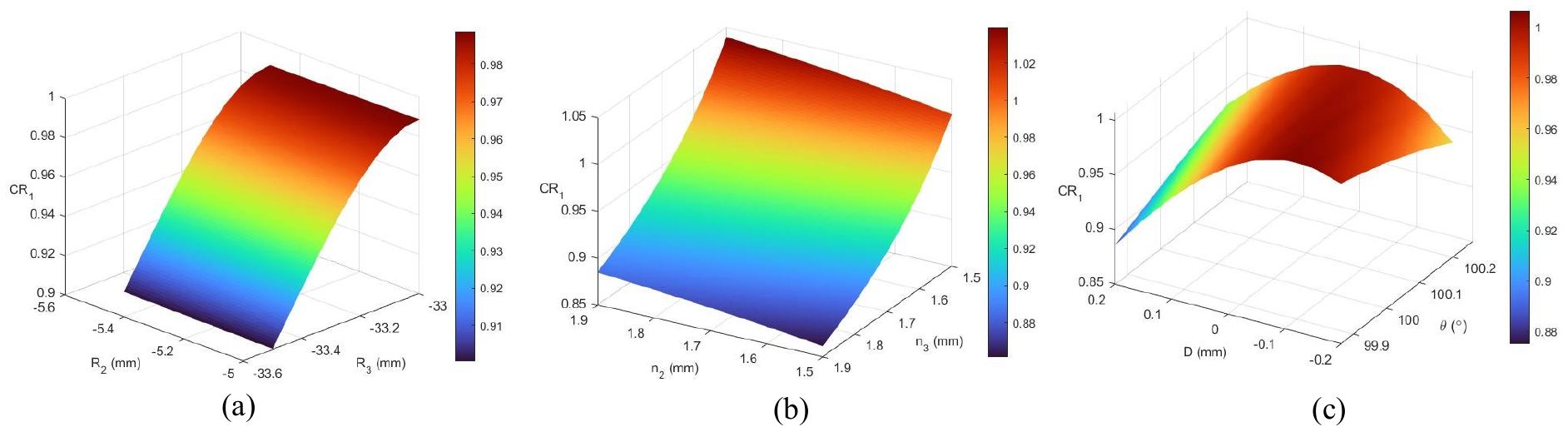}
\caption{Numerical relationship among PAL physical structure parameters and compactness ratio $CR_1$. (a) Curvature radii $R_2$, $R_3$, and compactness ratio $CR_1$. (b) Refractive index $n_{2}$, $n_{3}$, and compactness ratio $CR_1$. (c) Incidence angle $\theta$, Y-axis offset distance from initial incidence position $D$, and compactness ratio $CR_1$.}
\label{fig3}
\vskip-5ex
\end{figure*}

To analyze the impact of physical structural parameters on compactness more intuitively, we have iteratively calculated the curvature radii of the sensitive reflective surfaces $R_{2}$ and $R_{3}$ and obtained the compactness ratio through iterative ray tracing calculation. The numerical relationship is shown in Fig.~\ref{fig3}(a). Compared to the reflective surfaces $R_{2}$, $R_{3}$ is more sensitive to curvature changes. Thus, when designing compact PALs, more attention needs to be paid to the second reflective surface $R_{3}$. Due to the design goal of low-cost moldable or injectable PAL, it is considered to set the PAL block to a fixed low refractive index. The initial structure of the relay lens group consists of a cemented doublet used for correcting chromatic aberration and balancing the aberration of the PAL block. We studied the numerical relationship between the refractive index $n_{2}$ of the cemented doublet and the refractive index $n_{3}$ of the cemented doublet and the compactness ratio, as shown in Fig.~\ref{fig3}(b). When the positive lens uses a high refractive index, $n_{3}$ uses a low refractive index to achieve a better compactness ratio.
To further analyze the changes in the compactness ratio of PAL at the distance $D$ between the incident angle and the Y-axis direction variation range of the incident position. Due to a significant deviation in the Y-axis direction of the incident light or a significant change in the incident angle, it can cause significant ray aberration, resulting in total reflection or the inability of the light to trace to the image plane. Therefore, we set the range of distance variation and angle variation of the incident light in the Y-axis direction to a reasonable range and obtained the numerical relationship shown in Fig.~\ref{fig3}(c). On the one hand, the incident position moves to the negative direction of the Y-axis when the incident angle $\theta$ increases to greater than 100$^{\circ}$. On the other hand, the incident position moves to the positive direction of the Y-axis when the incident angle $\theta$ decreases to less than 100$^{\circ}$. Both of these cases result in a rapid decrease in compactness ratio.
Based on the above changes in physical parameters, the numerical analysis of the physical parameters' effect on the PAL compactness ratio can effectively guide us to design compact PAL physical structural parameters and further constrain their compactness ratio.
However, to design a PAL that can be applied to batch manufacturing, we need to consider the manufacturing challenges brought by more molding and injection molding technologies, as well as assembly processes.

\subsection{Consideration of manufacturing challenges for actual manufactured PALs}
For designing and processing high-performance PAL, there are more challenges related to manufacturing. Firstly, the edge thickness of PAL needs to be thick enough to meet the actual molding glass or injection molding process. Moreover, the edge thickness of PAL determines the clamping allowance after edge grinding. Although the PAL of F/\# 3.1 designed in  Fig.~\ref{fig1} has high imaging quality and small F/\#, its PAL block edge thickness is small. When considering increasing the margin of the effective aperture, the edge thickness will be smaller, making it impossible to achieve manufacturing and clamping. In addition, another manufacturing challenge is that the second reflecting surface and the first refracting surface belong to the spliced aspheric surface type, making the fabrication and measurement of spliced aspheric surfaces more difficult. To provide higher wear and corrosion resistance as the first lens, we plan to use molding glass technology to manufacture PAL blocks. To reduce the volume and weight of the relay lens, we use injection molding technology to manufacture the relay lens groups of PALs. Finally, due to the compactness of PAL lenses, strict control is also required for the manufacturing of PAL. Therefore, it is necessary for us to conduct a figure errors analysis on the entire PAL.

\subsection{Design, analysis, and manufacturing workflow of ASPAL}
\begin{figure*}[htbp]
\centering\includegraphics[width=13.2cm]{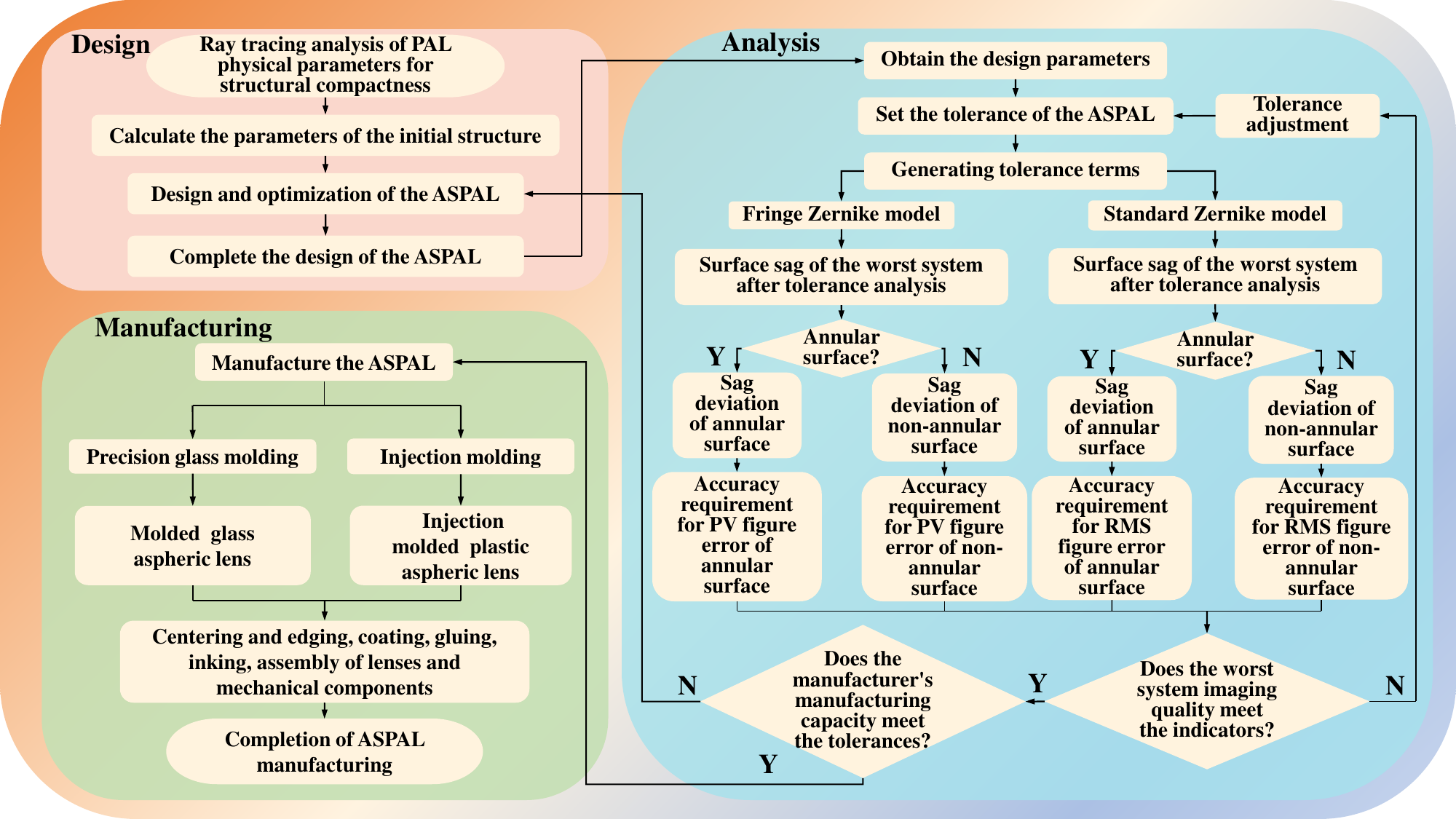}
\caption{Workflow of the high-performance glass-plastic hybrid minimalist ASPAL.}
\label{fig4}
\vskip-5ex
\end{figure*}

The workflow of the high-performance glass-plastic hybrid minimalist ASPAL is illustrated in Fig.~\ref{fig4}. The entire workflow consists of design, analysis, and manufacturing. Firstly, the ASPAL is designed based on the PAL imaging principle and the ray tracing model for the physical structure of the compact ASPAL (Section~\ref{sec:3}). Subsequently, to analyze the figure errors of annular aspheric surfaces and non-annular aspheric surfaces, we propose an aspheric surface tolerance analysis method (Section~\ref{sec:4}), which can effectively analyze the local tolerance of the ASPAL. Finally, based on the tolerance values obtained from the ASPAL tolerance analysis method, the ASPAL was fabricated using molded glass aspheric lenses and injection molded plastic aspheric lenses (Section~\ref{sec:5}).

\section{Design of the high-performance glass-plastic hybrid minimalist ASPAL}
\label{sec:3}
\subsection{Optical architecture design of ASPAL}
\label{sec:3.1}
In the design process, we first calculated the initial structure of ASPAL using the Petzval sum correction and focal power distribution theory, where the number of lenses is only 3~\cite{gao2022compact}. Based on the results of the ray tracing analysis in Section~\ref{sec:2.2}, we can effectively control and optimize the compactness of the PAL. Due to fewer initial structural optimization variables and the high-performance goal, all non-cemented surfaces are set as asphere surfaces, which can quickly optimize the design to obtain ASPAL designs close to the diffraction limit.
To facilitate the manufacturing and measurement of aspheric surfaces, we use even asphere surfaces, whose sag $z$ can be expressed as Eq.~(\ref{eq:9})~\cite{xue2018near}. In Eq.~(\ref{eq:9}), $c$ is the curvature, $r$ is the radius coordinate, $k$ is the conic coefficient and $\alpha$ is the $i$-th order aspheric coefficient.

\begin{equation}
\label{eq:9}
z=\frac{c r^2}{1+\sqrt{1-(1+k) c^2 r^2}}+\sum_{i=1}^8 \alpha_i r^{2 i}.
\end{equation}

\begin{figure}[ht!]
\centering\includegraphics[width=8.3cm]{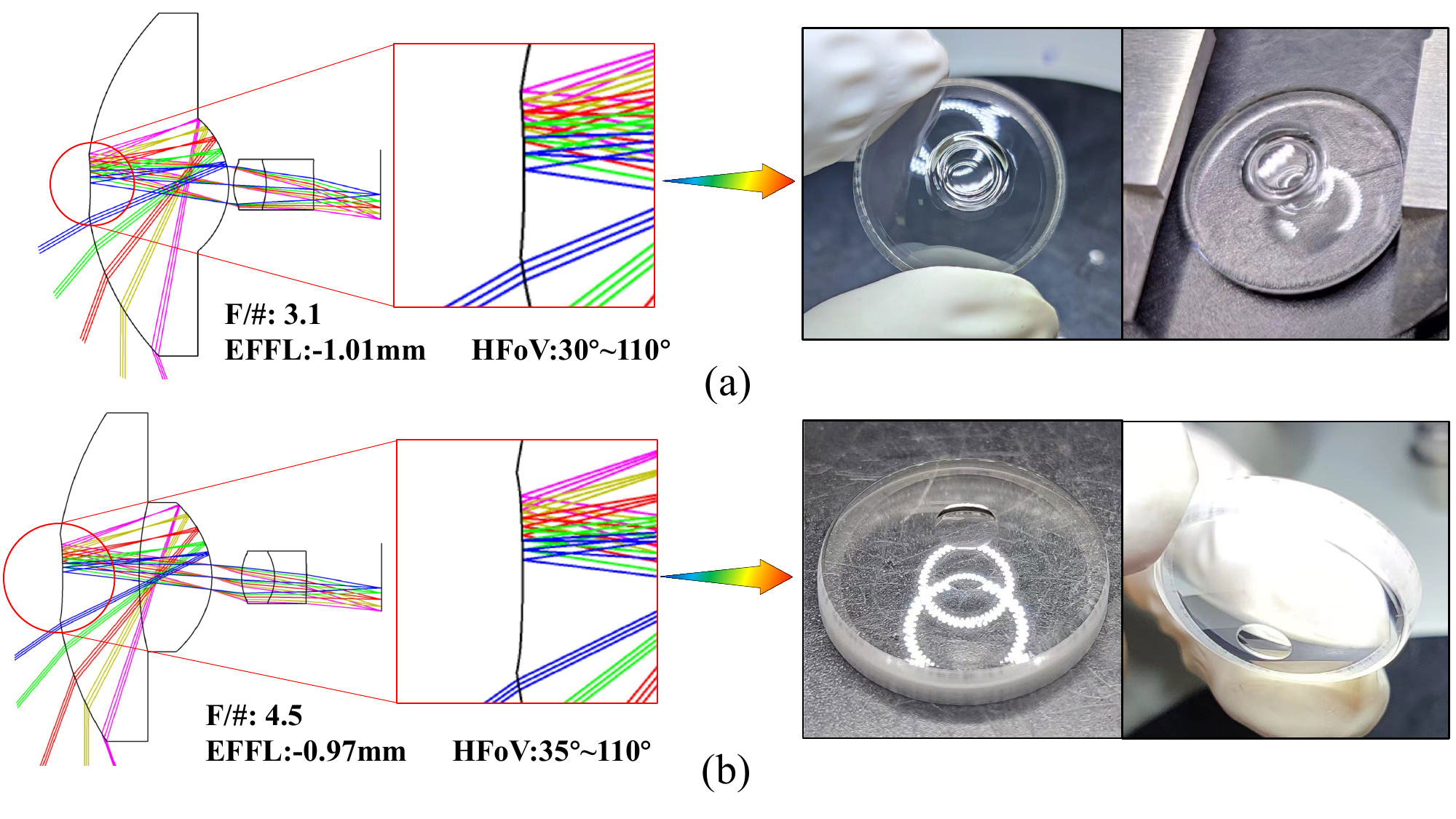}
\caption{Design compromise due to manufacturing challenges. (a) In the initial ASPAL design, due to the small reflection surface area, it was difficult to expel air from the lens during the molding process near the minimum FoV of 35°. (b) The improved design shifts the first transmission surface of ASPAL to the left, increases the area of the first reflection surface, and solves the difficulty of expelling air in the small FoV. However, the optical design indicators are compromised.}
\label{Fnumberchange}
\end{figure}

\begin{figure}[ht!]
\centering\includegraphics[width=8cm]{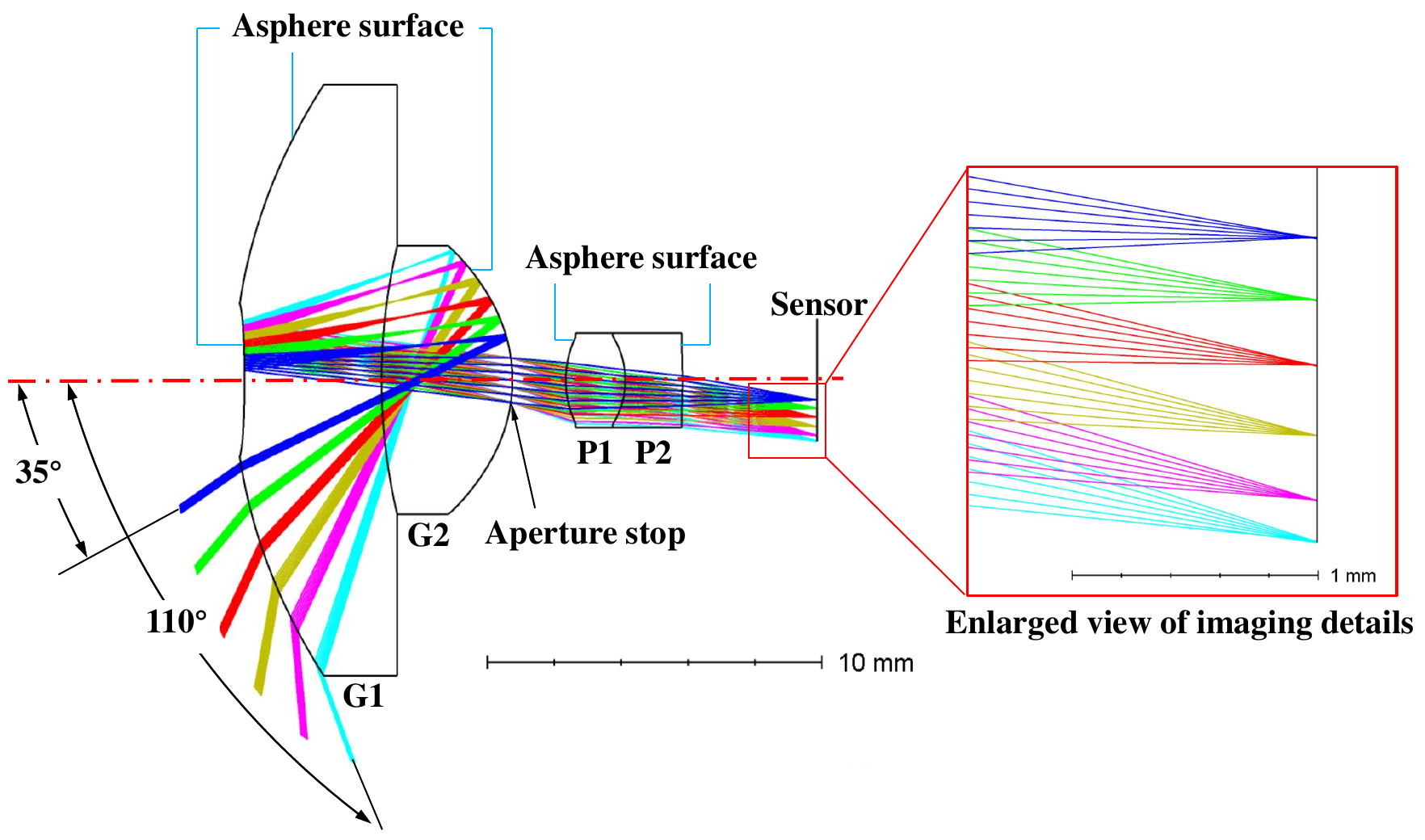}
\caption{Ultimate design of the high-performance glass-plastic hybrid minimalist PAL.}
\label{fig5}
\vskip-3ex
\end{figure}

Considering that glass has higher hardness and wear resistance compared to plastic, we set the first lens of ASPAL as a glass lens. To facilitate the mass production of ASPALs, we use molded glass technology to produce PAL blocks. Since the molded glass technology uses low-melting glass balls for mold pressing to obtain the final lens. Considering that the PAL block has a ladder on the outer diameter of the lens in the aperture, the difference in radius between the first transmission surface and the first reflection surface is significant. Thus, it is not conducive to molding and ensuring the accuracy of the lens surface shape. So we ultimately divide the PAL block into two low-melting glass lenses and mold them separately to ensure the accuracy of the aspherical surfaces of the molded glass lenses. Finally, they are glued together as a PAL block. To reduce the manufacturing cost of ASPAL, the two glass lenses G1 and G2 of the PAL block are made of the same glass material. Our initial design intention is to design a low-cost and high-performance ASPAL. To reduce the cost, the relay lens group is made of plastic material for P1 and P2 to facilitate mass manufacturing. The whole ASPAL consists of 4 lenses, with a reflective film coated on the front surface of G1 and an annular reflective film coated on the rear surface of G2 as an aperture stop. Due to the presence of the second reflective surface of the PAL block, there is a blind spot in the central area of the image formed by the PAL. In the initial ASPAL design, the FoV is $360^\circ{\times}(30^\circ{\sim}110^\circ$), and $F$$/$\# is 3.1, as shown in Fig.~\ref{Fnumberchange}(a). However, despite the initial designed ASPAL has a small blind spot, we encountered limitations during the experimental phase of mold manufacturing. Due to the small area of the second reflective surface, the first lens of the PAL block made of molded glass has the difficulty of expelling air at the edge of the reflecting surface. To address this manufacturing issue, we shift the transmission surface of the first glass lens to increase the reflecting surface area. Finally, the ASPAL glass mold fabrication for the first lens is achieved. The optical design compromise caused by manufacturing processes increases the difficulty of design and results in a decrease in performance. Consequently, considering the implementation of batch processing for mold manufacturing of glass, the minimum FoV of ASPAL was changed from $30^\circ$ to $35^\circ$, and $F$$/$\#  was increased from 3.1 to 4.5, as shown in Fig.~\ref{Fnumberchange}(b). The optical path of the ultimate ASPAL is shown in Fig.~\ref{fig5}.

\subsection{Imaging performance analysis}
The design parameters of the ASPAL are shown in Table \ref{tab:2}. We ultimately design a glass-plastic hybrid ASPAL with an $F$$/$\# of 4.5, a focal length of -0.97 mm, and a FoV of $360^\circ{\times}(35^\circ{\sim}110^\circ$). We chose a sensor with 3.75 $\upmu$m pixel size and 1280x960@60fps resolution. 

\begin{table}[!htbp]
\centering
\begin{scriptsize}
\caption{{Specifications of the ASPAL System}}
\begin{tabular}{cc} \hline
Parameters                   &    Specifications  \\ \hline
Wavelength                   &    486-656 nm\\
$F$$/$\#                   &    4.5  \\
Effective focal length       &    -0.97 mm \\
FoV                          &    360$^\circ$ $\times$(35$^\circ$ $\sim $110$^\circ$)\\
RMS spot radius                  &  2.1 $\upmu$m    \\
MTF at 133 lp/mm                 &  0.47         \\
Maximum $f$-$\theta$ distortion        &     3.5$\%$     \\
Minimum relative illumination        &     0.89  \\
Total length                 &   17.4 mm \\ \hline

\end{tabular}
\label{tab:2}
\end{scriptsize}
\vskip-2ex
\end{table}

\begin{figure}[ht!]
\centering\includegraphics[width=8.2cm]{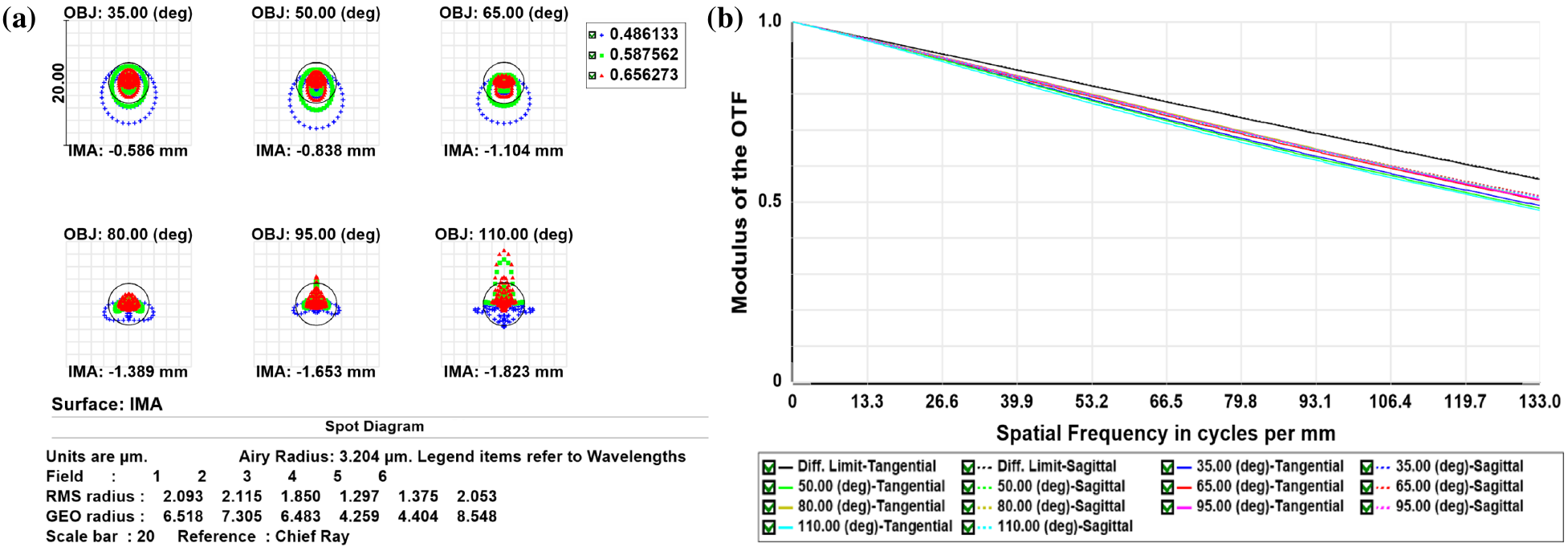}
\caption{The imaging quality of the ASPAL. (a) Spot diagram. (b) MTF.}
\label{fig6}
\vskip-3ex
\end{figure}

\begin{figure}[ht!]
\centering\includegraphics[width=8.2cm]{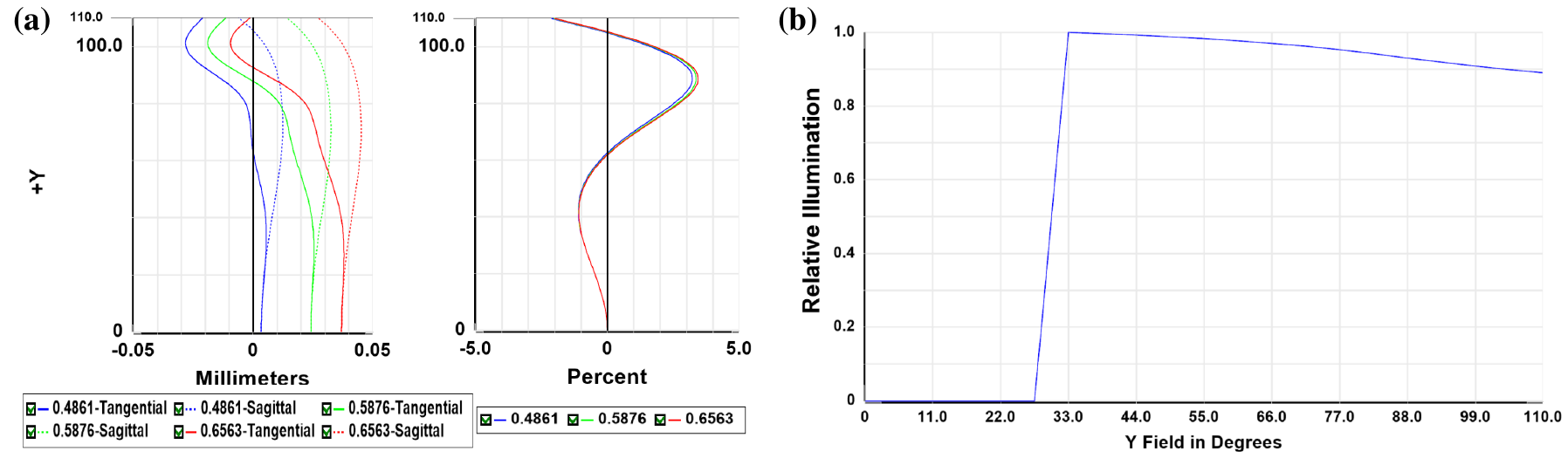}
\caption{The imaging quality of the ASPAL. (a) Field curvature and $f$-$\theta$ distortion. (b) Relative illumination.}
\label{fig7}
\vskip-3ex
\end{figure}

The maximum RMS spot radius of the ASPAL is 2.1 $\upmu$m, and the modulation transfer function (MTF) is higher than 0.47 at 133 lp/mm at the Nyquist frequency, which is close to the diffraction limit, as shown in Fig.~\ref{fig6}. Benefiting from high-order field curvature correction of aspheric surfaces, ASPAL's field curvature is less than 0.05 mm. The maximum $f-\theta$ distortion of ASPAL is approximately 3.5$\%$, and the minimum relative illumination is 0.89, as shown in Fig.~\ref{fig7}. The chief ray angle (CRA) of the selected sensor is linear 9$^\circ$, and the maximum CRA of the ASPAL design is 11.88$^\circ$. The CRA of the designed ASPAL should be as close as possible to the CRA of the sensor to reduce lens shading or color shading.

\section{High-performance ASPAL tolerance analysis methods}
\label{sec:4}
\subsection{Surface Irregularity Analysis}
\label{sec:4.1}
The manufacturing and measurement of aspheric surfaces play a crucial role in controlling the performance of the final designed ASPAL. The deviation between the actual sag of the surface of optical components manufactured and the designed sag is called sag deviation or figure error. To reasonably evaluate the local tolerance of ASPAL optical surfaces, we propose an ASPAL tolerance analysis method for the PV and RMS figure errors of annular and non-annular optical working surfaces. By using the Zernike Fringe Sag surface, a Fringe Zernike model can be constructed to evaluate the PV surface profile deviation of annular and non-annular surfaces. The Standard Zernike model can be constructed using the Zernike Standard Sag surface to evaluate the RMS figure error of annular and non-annular surfaces. The sag expression of the Zernike Fringe Sag surface is Eq.~(\ref{eq:10})~\cite{melich2013irregular}. The Zernike Fringe Sag surface is composed of the same polynomial as an even asphere surface and an aspheric term defined by an additional Zernike Fringe coefficient:

\begin{equation}
\label{eq:10}
z=\frac{c r^2}{1+\sqrt{1-(1+k) c^2 r^2}}+\sum_{i=1}^8 \alpha_i r^{2 i}+\sum_{i=1}^N A_i Z_i(\rho, \varphi).
\end{equation}

In Eq.~(\ref{eq:10}), $N$ is the number of Zernike coefficients in the order number, $A_i$ is the coefficient of the $i$-th Zernike Fringe polynomial, and $r$ is the radial ray coordinate. $\rho$ is the normalized radial ray coordinate, and $\varphi$ is the angular ray coordinate. The Zernike Fringe polynomial is defined by the Zernike Fringe Coefficients. The Zernike Fringe Sag surface has up to 37 Zernike Fringe polynomials. The Fringe Zernike model defined using the Zernike Fringe Sag surface can be used to evaluate small-amplitude random irregularity deviations on the surface. The surface sag deviation after tolerance analysis corresponds to the PV figure error of the optical surface.

The expression of the Zernike Standard Sag surface is similar to the Zernike Fringe Sag surface and both are expressed by Eq.~(\ref{eq:10}). The difference between the two surface types is that the Zernike Standard polynomial is defined by Zernike Standard Coefficients, and its Coefficients definition method is different from the Zernike Fringe polynomial. In the Zernike Standard Sag surface, $A_i$ are the coefficients of the $i$-th Zernike Standard polynomial. Their initial inputs are in the form of PV and RMS values that characterize the figure error, respectively. Therefore, we distinguish them and analyze them separately for tolerance. The Zernike Standard Sag surface has up to 231 Zernike Standard polynomials. Therefore, the Standard Zernike model constructed by the Zernike Standard Sag surface can evaluate the random irregularity deviation of small amplitudes on the surface. Since the Zernike Standard Sag surface has more Zernike Standard terms, more complex irregularity surfaces can be modeled. The surface sag deviation of this model after tolerance analysis corresponds to the RMS figure error of the optical surface.

 As shown in Fig. 4, we can derive the coefficients of the Zernike terms for each optical surface after tolerance analysis from optical design software. Then we can calculate the sag data of each surface after tolerance analysis using the Zernike Sag formulas. By using Matlab, the figure errors of aspheric surfaces after tolerance analysis can be calculated. For the PAL, this special optical system has both annular and non-annular surfaces. In tolerance analysis, it is necessary to eliminate invalid sag data in the central area of the annular surfaces. Finally, the RMS and PV values of the figure errors can be calculated by removing the invalid sag data from the center area of the annular surfaces.
Compared with traditional optical design software such as Zemax, our tolerance analysis method can realize tolerance analysis functions that cannot be realized by traditional optical design software which are essential for ASPAL tolerance analysis. Conventional optical design software is unable to realize the calculation of figure errors RMS and PV after tolerance analysis. Meanwhile, its sag data includes the sag of the center invalid region by default, so it is not possible to perform the tolerance analysis of the ASPAL systems with special annular surfaces at the effective optical aperture. By constructing the ASPAL tolerance analysis model, we could achieve the following new functions which cannot be realized by traditional optical design software.

(1) By calculating the figure errors of aspheric surfaces after tolerance analysis, the sag deviation data that can remove the data in the central region is obtained. These data can be used to eliminate invalid center area data based on the actual effective optical aperture. Thus, the sag data of each point in the tolerance analysis is completely within the effective aperture area of the annular and non-annular surfaces. The constructed tolerance analysis method of the aspheric surfaces is more conducive to the tolerance analysis of the ASPAL.

(2) By constructing the above ASPAL tolerance analysis model for annular and non-annular surfaces, the figure errors of annular and non-annular surfaces can be calculated. The PV and RMS values of the annular and non-annular surfaces can be obtained through the matrix data of figure errors.

After constructing the Fringe Zernike model using the Zernike Fringe Sag surface, we set the tolerance term for the annular and non-annular surfaces of ASPAL and analyze the PV figure error in the surface irregularity. Similarly, after we use the Zernike Standard Sag surface to construct the Standard Zernike model, we set the tolerance terms for the annular and non-annular surfaces of ASPAL and analyze the RMS figure error in the surface irregularity.

\begin{figure}[ht!]
\centering\includegraphics[width=8.2cm]{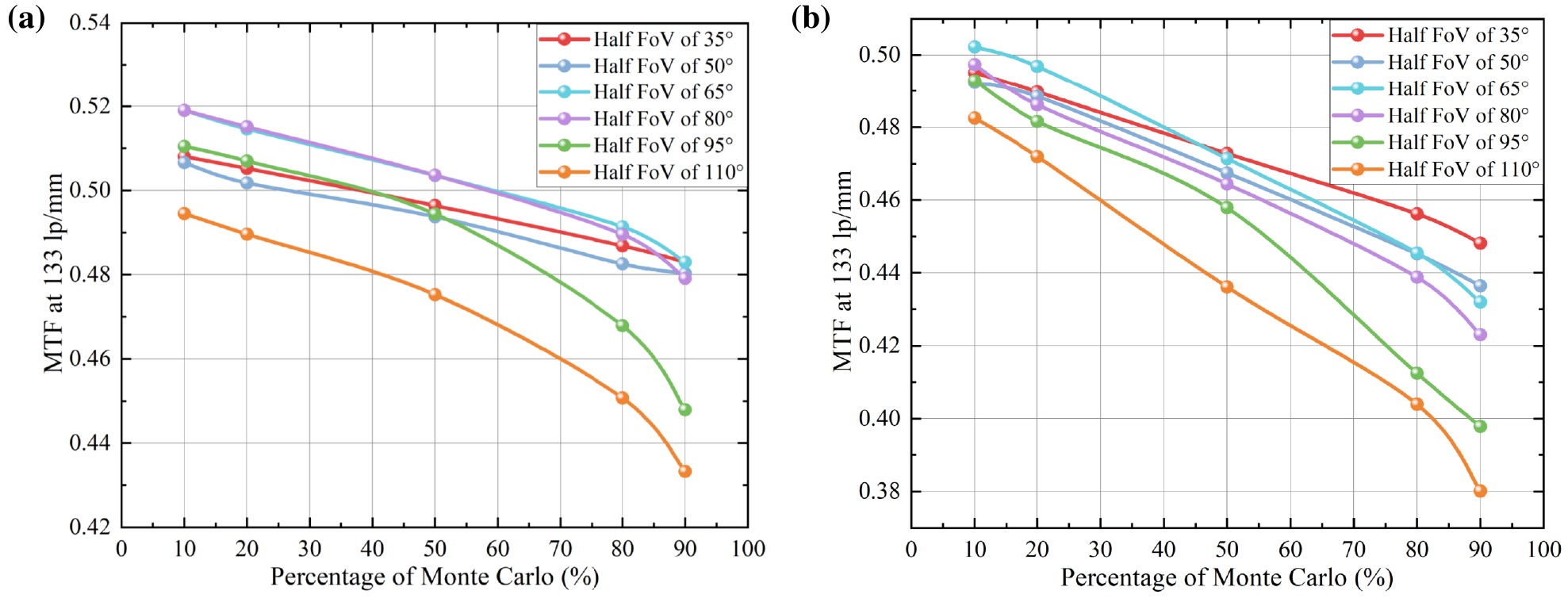}
\caption{MTF probability of ASPAL with the diffraction MTF average of 133lp$/$mm as the criterion. (a) Fringe Zernike model. (b) Standard Zernike model.}
\label{fig8}
\vskip-2ex
\end{figure}

 Finally, the MTF probability of ASPAL after 100 times of tolerance analysis using the diffraction MTF average of 133lp$/$mm as the criterion is shown in Fig.~\ref{fig8}.

\begin{figure}[ht!]
\centering\includegraphics[width=8.2cm]{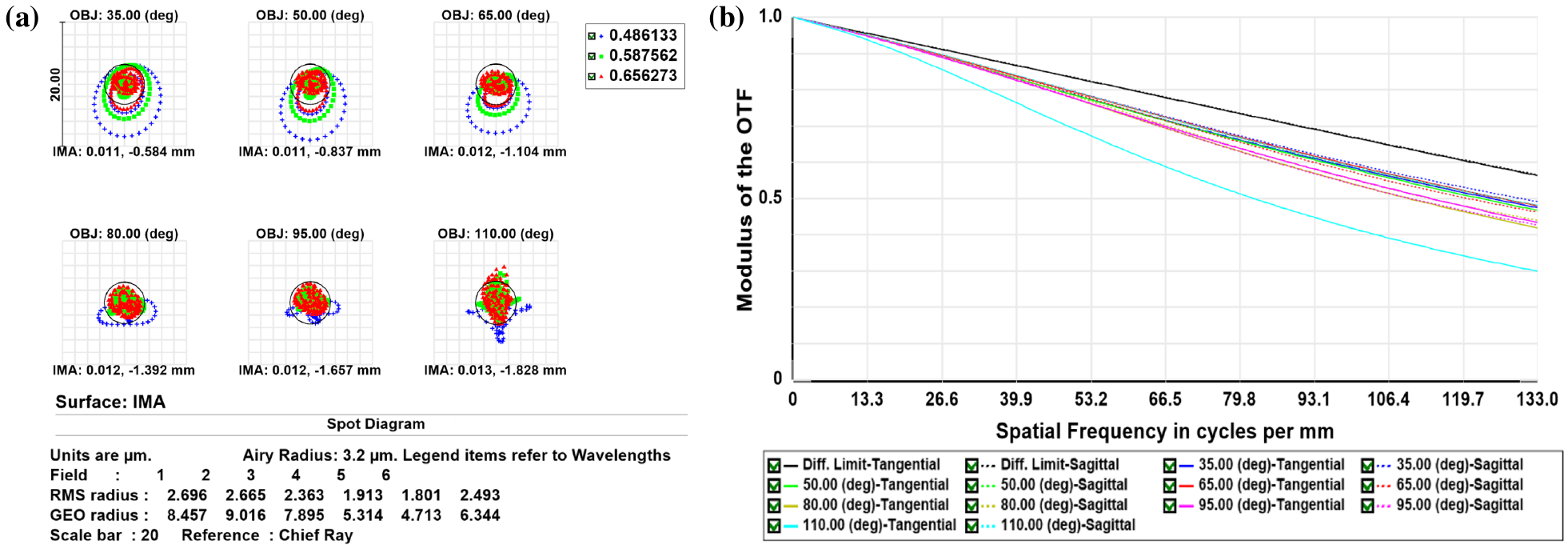}
\caption{Imaging quality of the worst ASPAL system with Fringe Zernike model.  (a) Spot diagram. (b) MTF.}
\label{fig9}
\vskip-3ex
\end{figure}

\begin{figure}[ht!]
\centering\includegraphics[width=8.2cm]{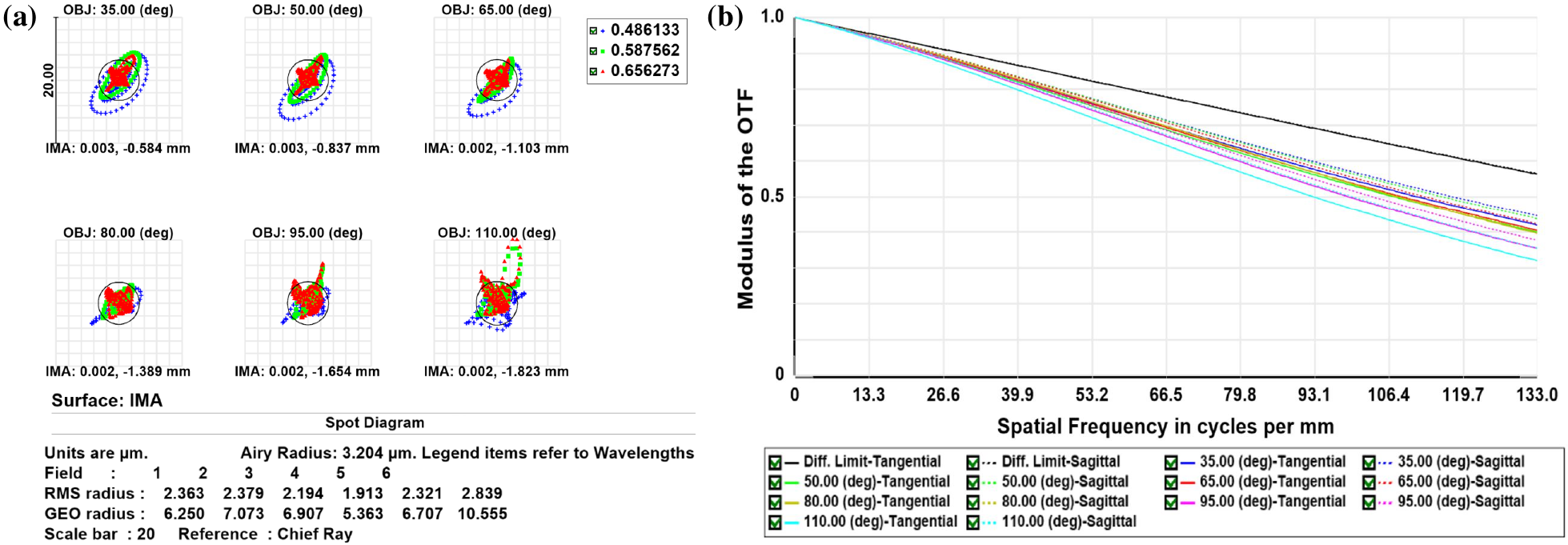}
\caption{Imaging quality of the worst ASPAL system with Standard Zernike model. (a) Spot diagram. (b) MTF.}
\label{fig10}
\vskip-3ex
\end{figure}

The MTFs obtained from 100 analyses using the Fringe Zernike model and the Standard Zernike model. They have a 90$\%$ probability of obtaining a full FoV with diffraction MTF averages higher than 0.41 and 0.38 at 133 lp$/$mm, respectively. The worst imaging quality of ASPAL obtained from the tolerance analysis of the Fringe Zernike model and Standard Zernike model are shown in Fig.~\ref{fig9} and Fig.~\ref{fig10}, respectively. Their RMS spot radii are less than 3.75 $\upmu$m pixel size. The MTFs are higher than 0.3 at 133 lp$/$mm.

\begin{figure}[ht!]
\centering\includegraphics[width=6.5cm]{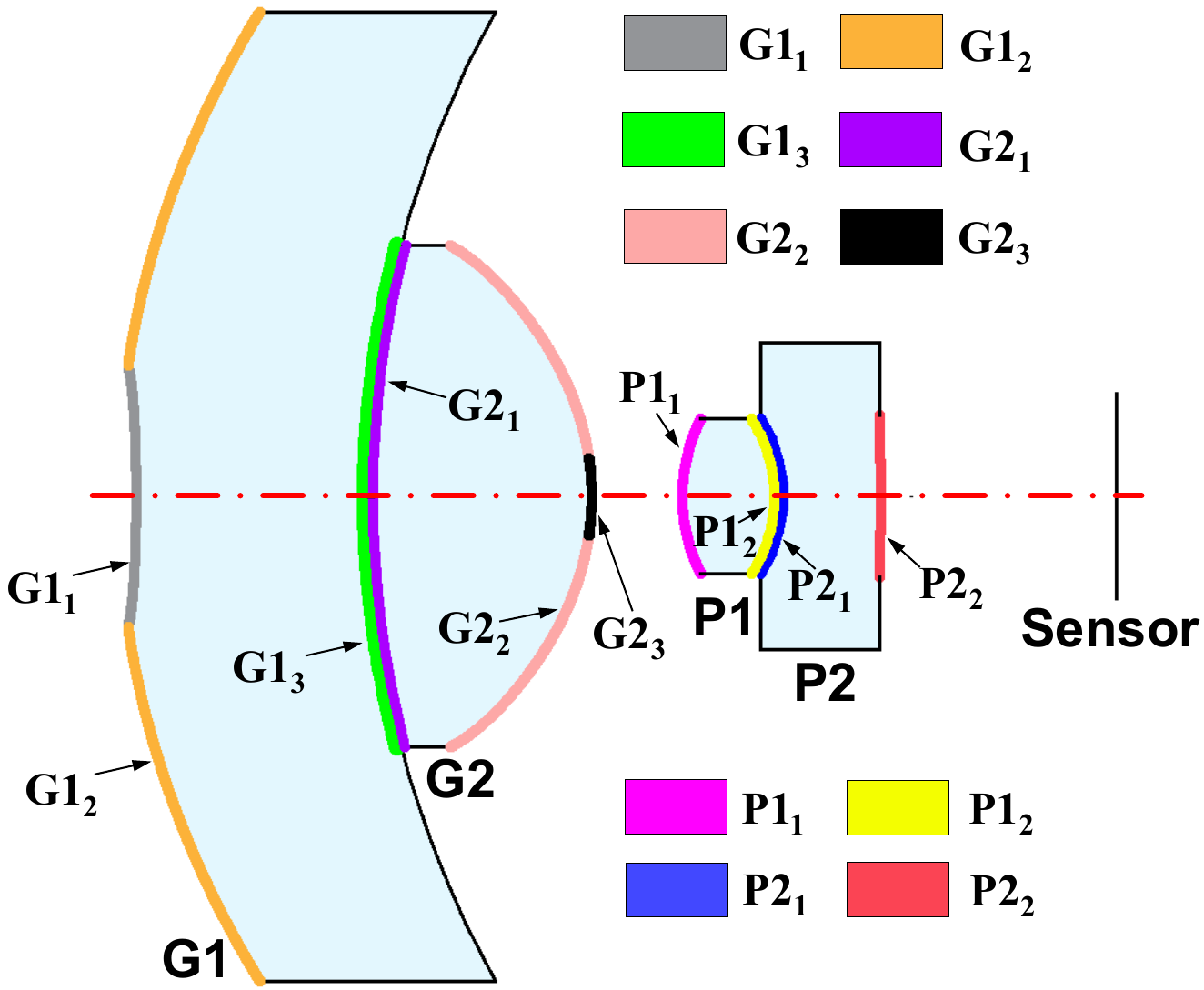}
\caption{Definition of surface names for the high-performance glass plastic hybrid ASPAL.}
\label{fig11}
\vskip-2ex
\end{figure}

Here, the Fringe Zernike model and the Standard Zernike model are not the final ASPAL surface irregularity tolerance terms, which need to be further calculated for the surface sag deviation. To facilitate the naming of each surface of ASPAL, the annular and non-annular surfaces on ASPAL of each surface in Fig.~\ref{fig11} are represented and named using different colors.

In the case that the surfaces are non-annular, the Fringe Zernike model and the Standard Zernike model can obtain the individual surface sag of the ASPAL system with the worst imaging quality. The ASPAL surface sag deviation $S_{x, y}$ can be calculated as the difference between the surface sag $S_{x, y}^t$ of the worst ASPAL and the corresponding surface sag $S_{x, y}^d$ of the designed ASPAL, as shown in Eq.~(\ref{eq:11}). The PV figure error $P V$ can be solved by the difference between the maximum value $\max\limits_{x, y} S_{x, y}$ and minimum value $\min\limits_{x, y} S_{x, y}$, as shown in Eq.~(\ref{eq:12}). Here, $x$ and $y$ represent the coordinate values of the surface sag in the X-axis and Y-axis directions. Although the Fringe Zernike model sets the PV figure error when setting the tolerance term, the RMS figure error calculated from the tolerance analysis can also be used as a reference value for the requirement of optical surface manufacturing. In the Standard Zernike model, the RMS deviation is set in the tolerance term, but the PV figure error of the deviation obtained from the tolerance analysis can also be used as the reference value for the manufacturing of optical surfaces. The RMS figure error $R M S$ is calculated as shown in Eq.~(\ref{eq:13}), with $M$ and $N$ representing the X-axis and Y-axis coordinates of the sampling points of each optical working surface.
The average value $\overline{S}$ of figure error can be calculated via Eq.~(\ref{eq:14}).

\begin{equation}
\label{eq:11}
S_{x, y}=S_{x, y}^t-S_{x, y}^d
{.}
\end{equation}

\begin{equation}
\label{eq:12}
P V=\max _{x, y} S_{x, y}-\min _{x, y} S_{x, y}
{.}
\end{equation}

\begin{equation}
\label{eq:13}
R M S=\sqrt{\frac{\sum_{x=1, y=1}^{M, N}\left(S_{x, y}-\overline{S}\right)^2}{M \times N}}
{.}
\end{equation}

\begin{equation}
\label{eq:14}
\overline{S}=\frac{\sum_{x=1, y=1}^{M, N} S_{x, y}}{M \times N}
{.}
\end{equation}

Since the ASPAL has annular surfaces, it is especially necessary to analyze its local tolerance. The optical surfaces of the annular surface G1$_2$ and G2$_2$ need to exclude the sag data of the non-optical working surfaces in the center. The figure error of the annular surface is the sag deviation $S_{x, y}^{\prime}$ between the sag $S_{x, y}^{\prime t}$ of the annular surface of the worst imaging quality ASPAL after tolerance and the surface sag $S_{x, y}^{\prime d}$ of the design value, as shown in Eq.~(\ref{eq:15}).

\begin{equation}
\label{eq:15}
S_{x, y}^{\prime}=S_{x, y}^{\prime t}-S_{x, y}^{\prime d}
{.}
\end{equation}

\begin{equation}
\label{eq:16}
P V_a=\max _{x, y} S_{x, y}^{\prime}-\min _{x, y} S_{x, y}^{\prime}
{.}
\end{equation}

\begin{equation}
\label{eq:17}
R M S_a=\sqrt{\frac{\sum_{x=1, y=1}^{M, N}\left(S_{x, y}^{\prime}-\overline{S^{\prime}}\right)^2}{M \times N}}
{.}
\end{equation}

\begin{equation}
\label{eq:18}
\overline{S^{\prime}}=\frac{\sum_{x=1, y=1}^{M, N} S_{x, y}^{\prime}}{M \times N}
{.}
\end{equation}

The difference between the maximum $\max\limits_{x, y} S_{x, y}^{\prime}$ and minimum $\min\limits_{x, y} S_{x, y}^{\prime}$ surface sag deviation of the annular surface is the PV figure error $P V_a$ of the annular surface, which can be calculated as Eq.~(\ref{eq:16}). The RMS figure error $R M S_a$ of the annular surface can be solved by Eq.~(\ref{eq:17}). The average value of the annular surface can be calculated by Eq.~(\ref{eq:18}). In this case, the calculated $P V_a$ and $R M S_a$ figure errors of the annular surfaces are the final local tolerance requirements. This method of surface irregularity tolerance analysis of the annulus surfaces can be used not only in ASPAL but also in optical systems with central obscuration, such as the Cassegrain optical system, for evaluating local tolerances.

\subsection{Consideration of mid-spatial frequency surface errors for ASPAL}
For diamond turning of aspheric molds, it is common to use compliant sub-aperture tools, which may cause mid-spatial frequency surface errors on the molds. These mid-spatial frequency surface errors can lead to reduced system resolution, stray light, reduced uniformity of the illumination system, and so on~\cite{harvey1995scattering}. Therefore the influence of mid-spatial frequency surface errors on the ASPALs needs to be considered besides the surface irregularities characterized by PV and RMS~\cite{youngworth2000simple,kumler2007measuring,aikens2008specification,tamkin2010theory,murphy2015considerations,messelink2018mid}.

In the initial tolerance analysis process, we also considered the possible mid-spatial frequency errors caused by single point diamond turning of the mold. To evaluate the low and mid-spatial frequency surface errors caused by periodic ripple trajectories generated during the single point diamond manufacturing mold process, the low and mid-spatial frequency surface can be used for tolerance analysis. Its sag can be characterized by Eq.~(\ref{eq:199}), which essentially takes the form of adding a periodic surface on the Zernike Standard Sag surface.

\vskip-3ex
\begin{equation}
\label{eq:199}
\begin{split}
&z=\frac{c r^2}{1+\sqrt{1-(1+k) c^2 r^2}}+\sum_{i=1}^8 \alpha_i r^{2 i}+\sum_{i=1}^N A_i Z_i(\rho, \varphi) \\
&-\frac{A}{2}\left[\cos \left(2 \pi \omega_0 r+\varphi_0\right)-\cos \varphi_0\right].
\end{split}
\end{equation}
\vskip-1.5ex

In Eq. (\ref{eq:199}), its expression is the sum of the expressions for the Even Asphere part, Zernike Standard Sag part, and Periodic Sag part. The periodic part is a sag value superimposed on the surface with a fixed amplitude and frequency. For the expression of the periodic sag component, ${A}$ is the amplitude of the periodic term, $\omega_0$ is the frequency of the periodic term, and $\varphi_0$ is the phase offset. For the ripple phenomenon generated by this periodic surface, the above parameters can be used to control its ripple period, amplitude, and phase offset.

\begin{figure}[ht!]
\centering\includegraphics[width=8cm]{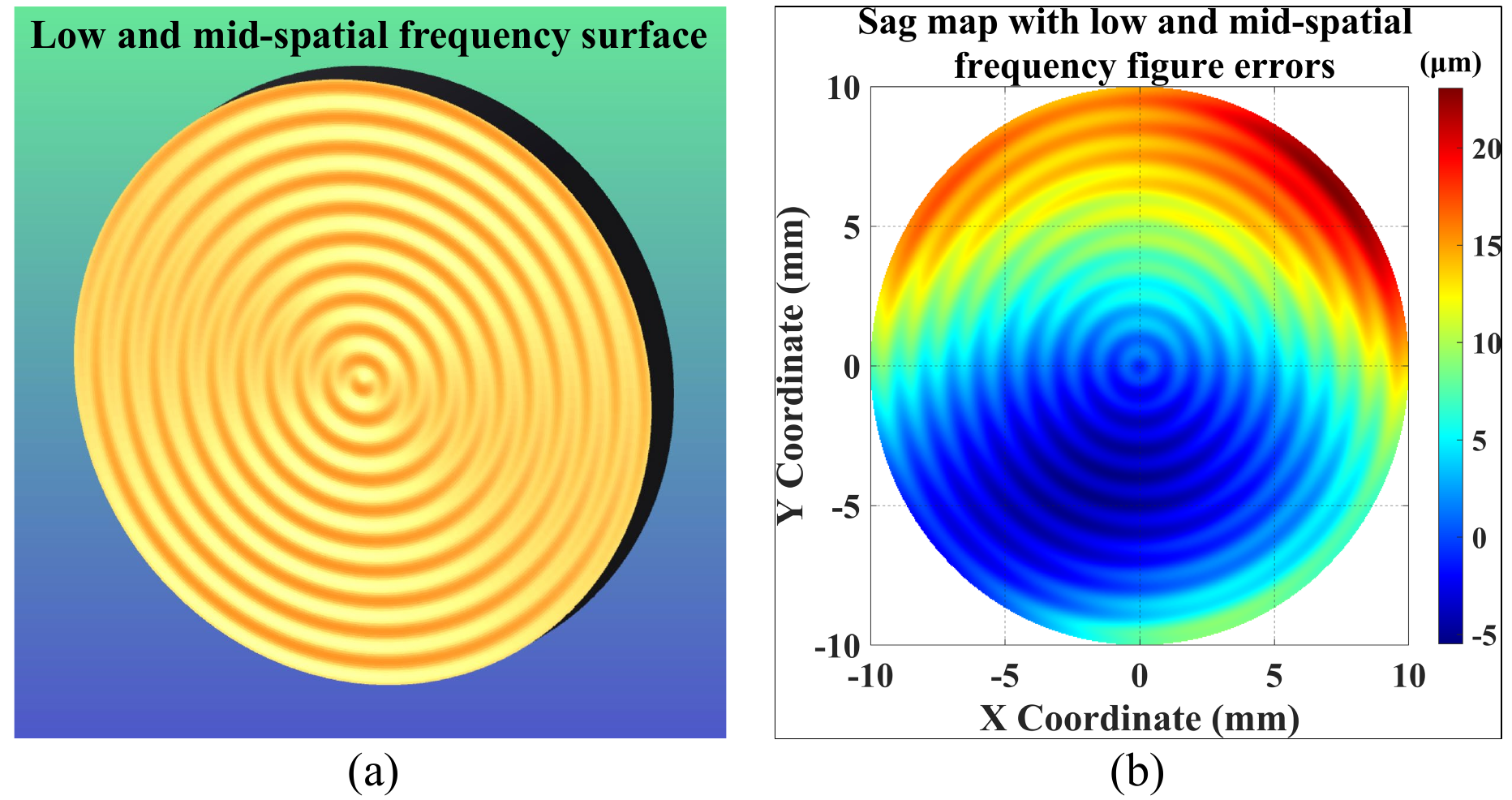}
\caption{An optical surface characterized by surface irregularities using low and mid-spatial frequency surface. (a) 3D view. (b) Sag map.}
\label{ripple}
\vskip-1.5ex
\end{figure}

\begin{figure}[ht!]
\centering\includegraphics[width=8cm]{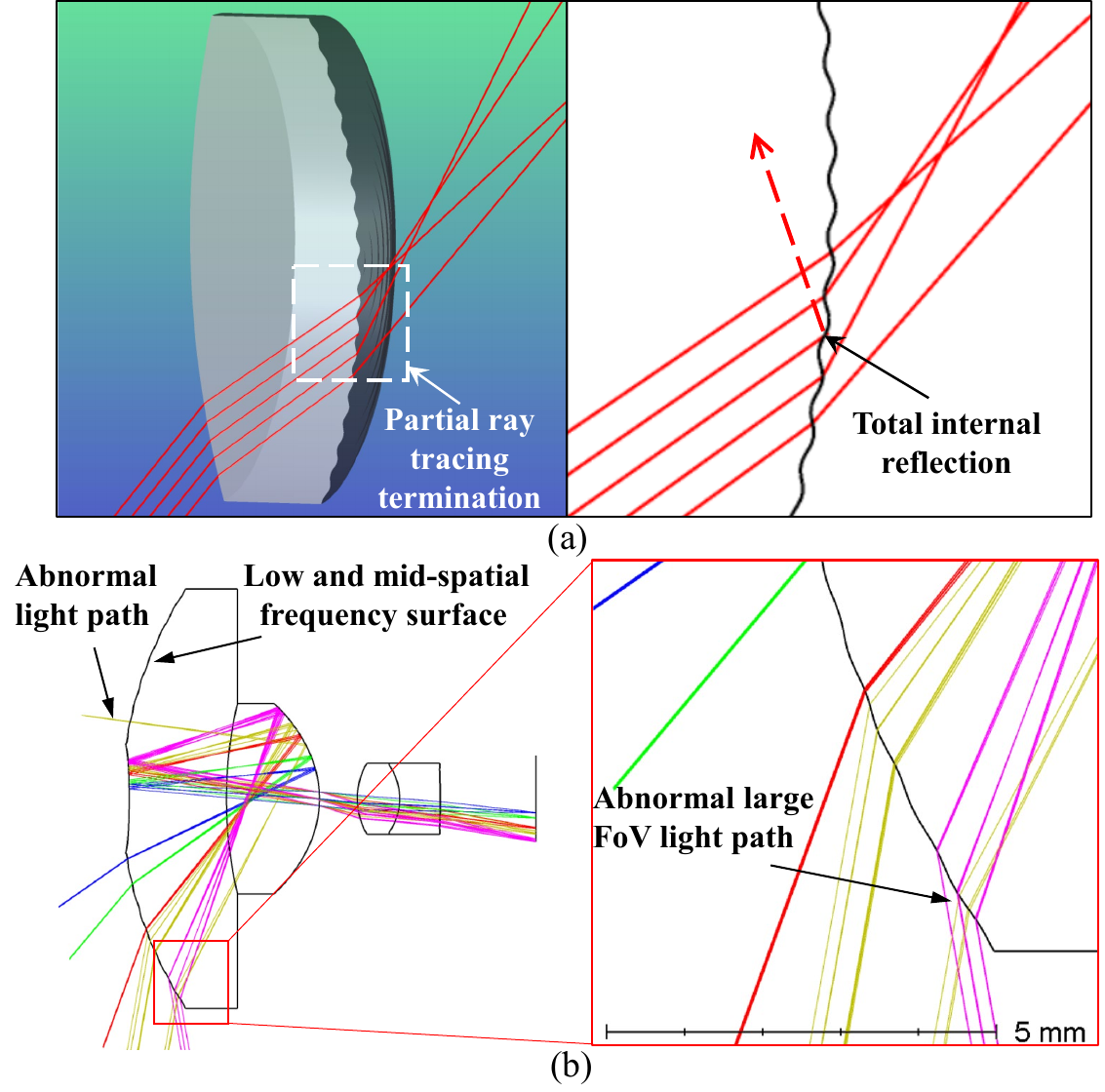}
\caption{Total internal reflection (TIR) phenomenon on ripple surfaces. (a) TIR occurs in the large FoV rays. (b) Large FoV abnormal rays in ASPAL's G1$_2$ after applying the low and mid-spatial frequency surface.}
\label{TIR}
\end{figure}

In Fig.~\ref{ripple}, we present a 3D view and sag map of an even asphere surface with irregularity tolerance using a low and mid-spatial frequency surface. After obtaining the sag map, we can use the same analysis process to perform tolerance analysis on the surface irregularity in Section~\ref{sec:4.1}.
This low and mid-spatial frequency surface with periodic surface properties can be used to evaluate the effect of mid-frequency errors on optical aspheric surfaces generated by single-point diamond turning molds. However, for incident rays at a large FoV, total internal reflection (TIR) usually occurs on the surface of the ripple surface, as shown in Fig.~\ref{TIR}(a). On its local enlarged image, some rays cannot continue to be traced due to TIR. We have drawn the TIR path of the abnormal rays. Therefore, large FoV incident rays on the ripple surface can lead to possible TIR, ultimately resulting in inaccurate calculation results. For ASPAL systems with extremely large FoV, we have also attempted for the first time to use low and mid-spatial frequency surfaces on the ASPAL’s G1$_2$, as shown in Fig.~\ref{TIR}(b). The abnormal optical path with a large FoV is different from the imaging optical path, which can cause abnormal ray tracing and even lead to ray tracing collapse. However, the current illustration only shows the first case where a low and mid-spatial frequency surface is used. Current results show that if the ripple surface is applied to multiple aspheric surfaces in ASPAL, more TIR phenomena will occur in large FoV rays, making the system unable to model. Compared to large aperture optical systems, the imaging quality of small aperture optical systems is less affected by mid-spatial frequency surface errors with a given sub-aperture tool used in asphere manufacturing. In addition, mid-spatial frequency errors have less effect on larger $F$$/$\# optical systems than smaller $F$$/$\#  optical systems~\cite{adams2018understanding}. As discussed in Section~\ref{sec:3.1}, the $F$$/$\#  for the final ASPAL design is increased from 3.1 to 4.5 due to the difficulty of expelling air in small reflective surface  G1$_1$. Therefore, the $F$$/$\#  of the final ASPAL and small diameter ASPAL will make the ASPAL less affected by the mid-spatial frequency surface errors. Based on the above reasons, we use the Zernike Sag surfaces in Section~\ref{sec:4.1} for tolerance analysis.

\subsection{Tolerance analysis results}
\label{sec:4.3}
\begin{figure*}[ht!]
\centering\includegraphics[width=15cm]{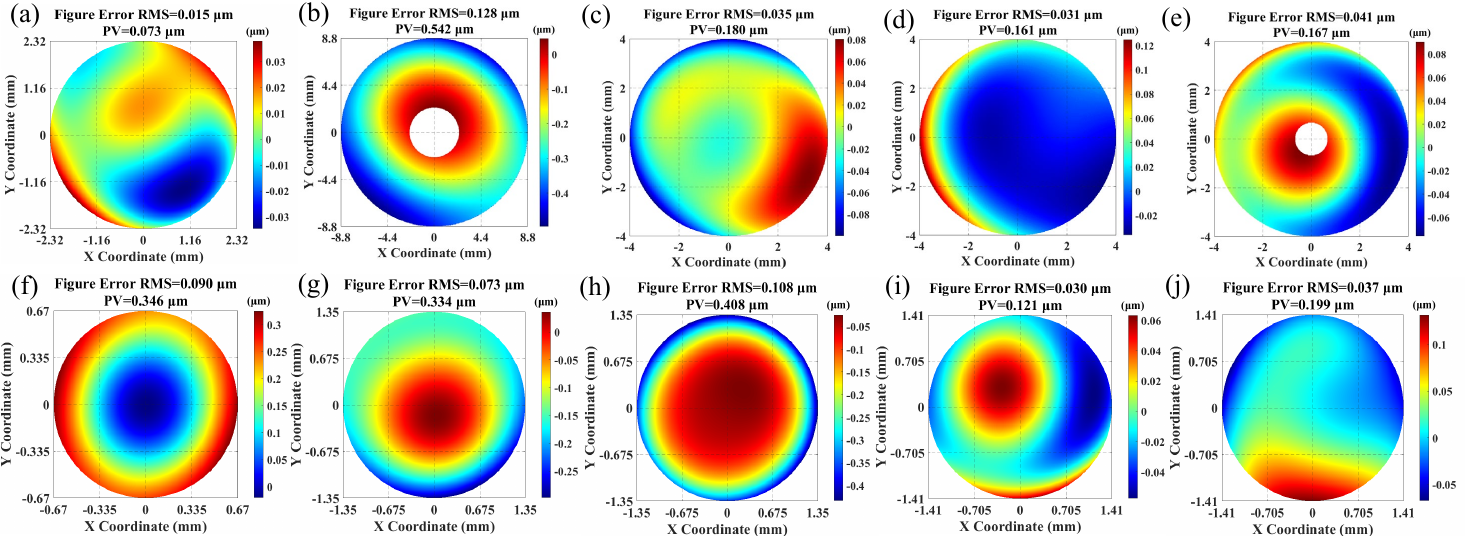}
\vskip-2ex
\caption{Figure errors of the worst ASPAL system with Fringe Zernike model. (a) G1$_1$. (b) G1$_2$. (c) G1$_3$. (d) G2$_1$. (e) G2$_2$. (f) G2$_3$. (g) P1$_1$. (h) P1$_2$. (i) P2$_1$. (j) P2$_2$.}
\label{fig12}
\vskip-3ex
\end{figure*}

\begin{figure*}[ht!]
\centering\includegraphics[width=15cm]{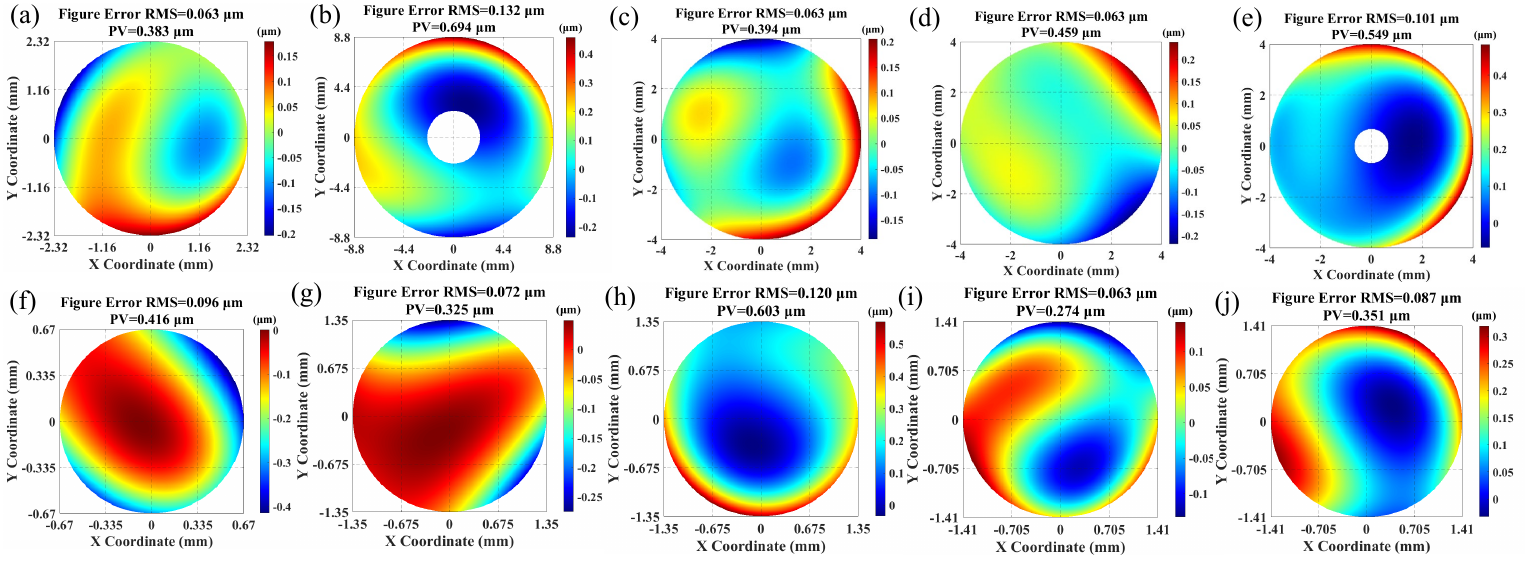}
\vskip-2ex
\caption{Figure errors of the worst ASPAL system with Standard Zernike model. (a) G1$_1$. (b) G1$_2$. (c) G1$_3$. (d) G2$_1$. (e) G2$_2$. (f) G2$_3$. (g) P1$_1$. (h) P1$_2$. (i) P2$_1$. (j) P2$_2$.}
\label{fig13}
\vskip-5ex
\end{figure*}

The calculated figure errors PV, and RMS value of each surface of ASPALs after analyzing the Fringe Zernike model and Standard Zernike model with the worst imaging quality systems are shown in Fig.~\ref{fig12} and Fig.~\ref{fig13}, respectively. The optical surface sag sampling point is 513$\times$513, and its center point coordinate is 257$\times$257. It should be noted in the calculation process that we only select the sampling point of the optical working surfaces on the annular surfaces and the non-annular surfaces.

Compared to the Fringe Zernike model tolerance analysis results, the Standard Zernike model tolerance analysis resulted in figure errors with more complex surface irregularities. Although the PV of each surface analyzed by both tolerance analysis methods is within 1 $\upmu$m, the RMS figure error is within 0.13 $\upmu$m. However, because the Fringe Zernike model contains fewer Zernike terms, the simulation of surface irregularities is not as robust as the Standard Zernike model. For example, the PV value of surface G1$_2$ is significantly smaller compared to other optical surfaces, and the Standard Zernike model is more recommended because it has more Zernike terms. Therefore, we use the PV and RMS values obtained from the Standard Zernike model analysis as the fabrication requirements for each optical surface.

Combining the results of the two model tolerance analyses, we determine the tolerance values for the final manufactured ASPAL as shown in Table~\ref{tab:3}.

\begin{table}[!htbp]
\centering
\begin{scriptsize}
\caption{{Tolerance values of the ASPAL system}}
\begin{tabular}{cc} \hline
Tolerance items                  &    Value  \\ \hline
Radius (fringe)                  &   $ \leqslant3  $\\
Thickness (mm)                   &    ±0.01  \\
Surface decenter ($\upmu$m)            &    ±5 for PAL block\\  
                                 &    ±4 for Relay lens\\
Surface tilt ($^{\prime}$)         &    ±1  \\
Element decenter ($\upmu$m)            &    ±5 for PAL block\\  
                                 &    ±4 for relay lens\\
Element tilt ($^{\prime}$)         &    ±3 for PAL block\\  
                                 &    ±1 for Relay lens\\
Surface irregularity ($\upmu$m)    &   See Section~\ref{sec:4.3}\\
Refractive index                 &    ±0.001 \\
Abbe number (\%)                 &    ±1 \\ \hline

\end{tabular}
\label{tab:3}
\end{scriptsize}
\vskip-2ex
\end{table}

\section{Manufacturing and testing of high-Performance glass-plastic hybrid minimalist ASPAL}
\label{sec:5}
\subsection{ASPAL batch manufacturing based on precision glass molding and injection molding}
\begin{figure*}[ht!]
\centering\includegraphics[width=12cm]{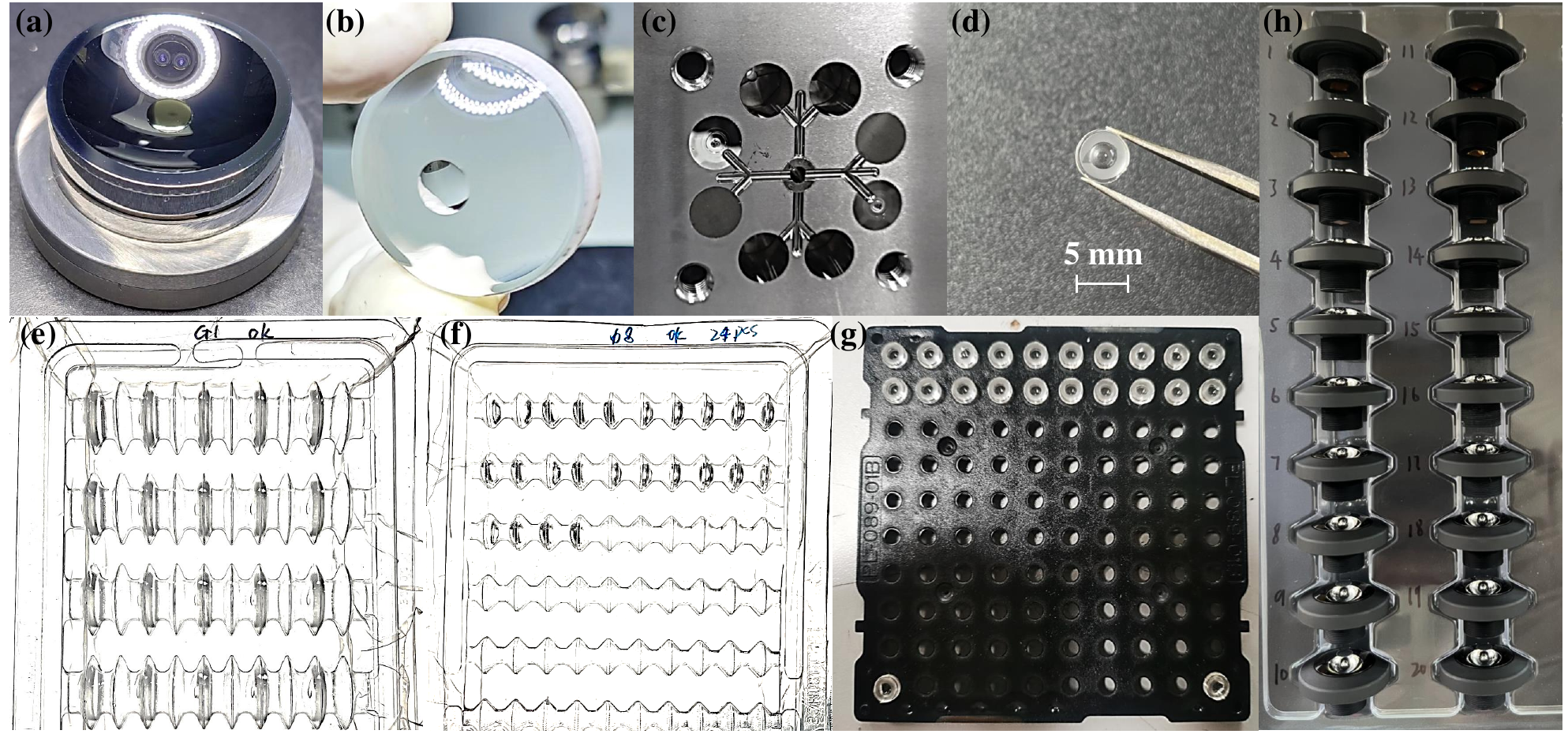}
\caption{ASPAL batch manufacturing based on molded glass technology and injection molding technology. (a) G1 aspheric mold. (b) Aspheric lens G1 manufactured by glass molding technology. (c) Plastic aspheric mold. (d) Aspheric lens P2 manufactured by injection molding. (e) G1 manufactured in small batches. (f) G2 manufactured in small batches. (g) P1 and P2 cemented lens groups. (h) 20 high-performance glass-plastic hybrid ASPALs.}
\label{fig14}
\vskip-4ex
\end{figure*}

Aspheric lenses can be fabricated quickly and cost-effectively using glass molding technology and injection molding aspheric technology. Fig.~\ref{fig14} shows an aspherical mold and an injection mold, G1 and G2 manufactured in bulk by glass molding technology, injection molding technology, cemented len groups P1 and P2, and 20 ASPALs assembled in bulk.

\subsection{Manufacturing of ASPAL and comparison between ASPAL and traditional SPALs}
\begin{figure}[ht!]
\vskip-2ex
\centering\includegraphics[width=7cm]{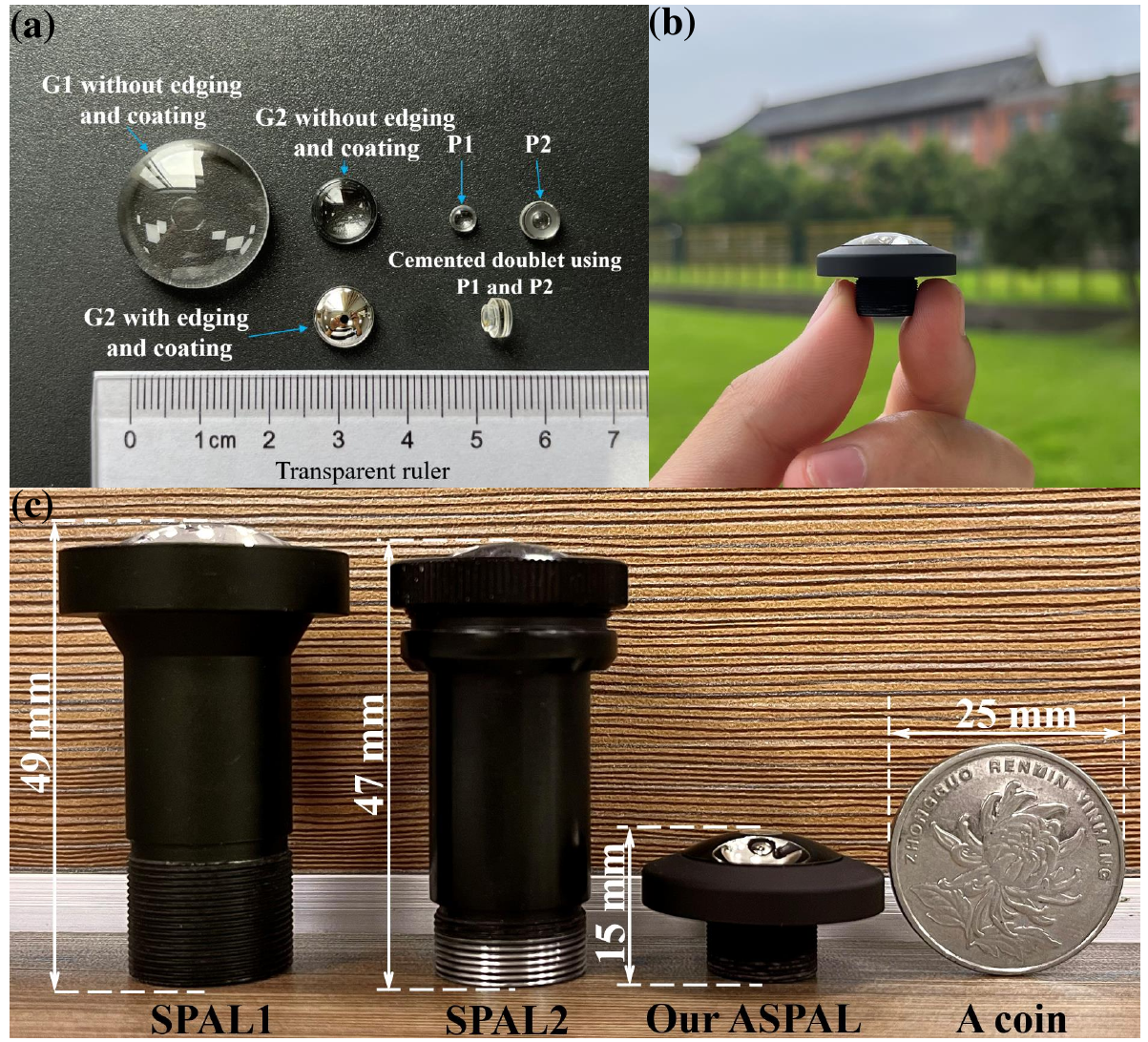}
\caption{Manufacturing of ASPAL and comparison between the ASPAL and traditional SPALs. (a) Lens components of the ASPAL. (b) ASPAL prototype. (c) Comparison of dimensions among our ASPAL, traditional SPALs, and a coin.}
\label{fig15}
\vskip-5ex
\end{figure}

A single ASPAL consists of G1, G2, P1, and P2 lenses as shown in Fig.~\ref{fig15}~(a). The small reflective surface of G1 is coated with a reflective film, and the rear surface of G2 is coated with an annular reflective film to serve as an aperture stop. The coated G1 and G2 lenses are cemented together as a PAL block. P1 and P2 are cemented together as a relay lens group. The assembled ASPAL prototype is shown in Fig.~\ref{fig15}~(b). Compared with the previously fabricated conventional spheric panoramic annular lenses (SPALs), our proposed ASPAL increases the full FoV by more than 30$^{\circ}$, reduces the lens height by more than 68$\%$, and reduces the lens weight by about 80$\%$, as shown in Fig.~\ref{fig15}~(c). A comparison of the detailed parameters of SPALs and our ASPAL is shown in Table~\ref{tab:Comparison}.

\begin{table*}[!ht]
\centering
\scriptsize
\caption{Comparison of parameters between high-performance glass plastic hybrid minimalist ASPAL and traditional SPALs.}
\setlength{\tabcolsep}{1.5mm}{
\begin{tabular}{cccc}
\toprule
\makecell[c]{Specification}&\makecell[c]{SPAL1}& \makecell[c] {SPAL2}&\makecell[c]{ASPAL}\\
\midrule
\makecell[c]{Sensor size (inch)}&\makecell[c]{1/3}& \makecell[c]{1/3}&\makecell[c]{1/3}\\
\midrule
\makecell[c]{FoV ($^\circ$)}&\makecell[c]{360$^\circ$ $\times$(30$^\circ$ $\sim $90$^\circ$)}& \makecell[c] {360$^\circ$ $\times$(30$^\circ$ $\sim $85$^\circ$)}&\makecell[c]{\textbf{360$^\circ$ $\times$(35$^\circ$ $\sim $110$^\circ$)}}\\
\midrule
\makecell[c]{Height (mm)}&\makecell[c]{49}&\makecell[c] {47}&\makecell[c]{\textbf{15}}\\
 \midrule
\makecell[c]{Maximum diameter (mm)}&\makecell[c]{31}& \makecell[c]{26}&\makecell[c]{26}\\
\midrule
\makecell[c]{Weight (g)}&\makecell[c]{42.7}& \makecell[c] {39.1}&\makecell[c]{\textbf{8.5}}\\

\bottomrule
\label{tab:Comparison}
\end{tabular}}
\vskip-3ex
\end{table*}

\subsection{Figure errors test and imaging performance test of ASPAL}
\begin{figure}[ht!]
\centering\includegraphics[width=8cm]{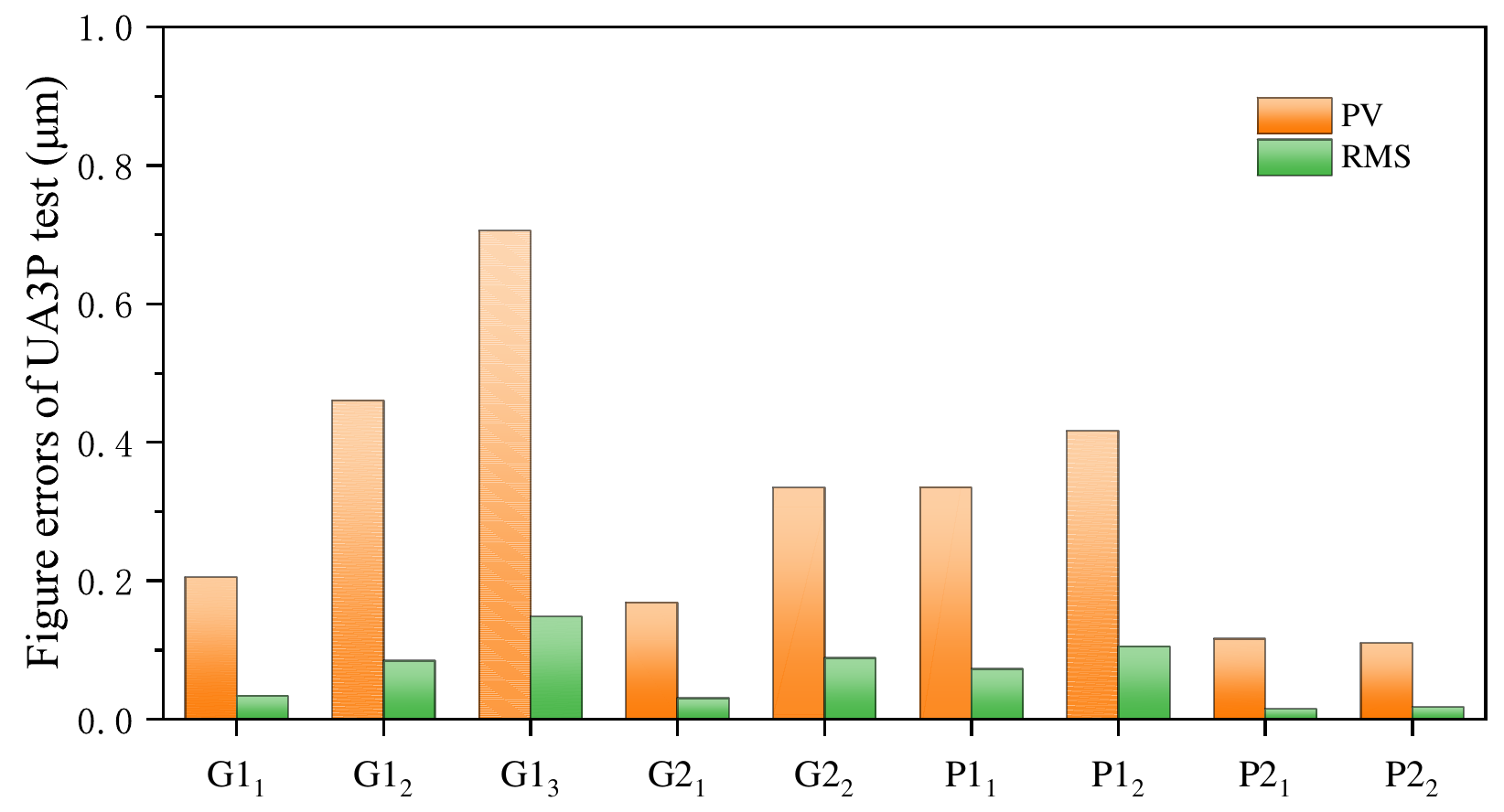}
\caption{Figure errors of manufactured the ASPAL obtained from optical profilometer Panasonic UA3P test.}
\label{fig16}
\vskip-2ex
\end{figure}

Figure errors test of aspheric surfaces is essential to the ASPAL imaging performance. We use a high-precision optical profilometer UA3P from Panasonic Corp. to measure the figure errors of each surface of the ASPAL. The test results are shown in Fig.~\ref{fig16}. Except for the large figure errors of G1$_3$, the third surface of the first molded lens with a large aperture, the figure errors of the other surfaces are within about RMS 0.1 $\upmu$m and PV 0.5 $\upmu$m. The reason for this is that we tested the figure errors of G1$_3$ with a diameter of 17 mm in full aperture, and therefore its figure errors are larger. As the effective aperture of the face type G1$_3$ is 8 mm, its figure errors will be less than the test value. As for the injection-molded lens P1, its P1$_2$ RMS is close to 0.1 $\upmu$m and its PV is about 0.4 $\upmu$m. Although it is larger than the previous tolerance analysis results, the figure errors can be further reduced by several corrections of the mold. By multiple measuring the surface sag of aspheric optical components after glass molding and injection molding, the sag of the aspheric molds can be further iteratively corrected, thereby obtaining better surface shapes of aspheric molds.

\begin{figure}[ht!]
\centering\includegraphics[width=8.2cm]{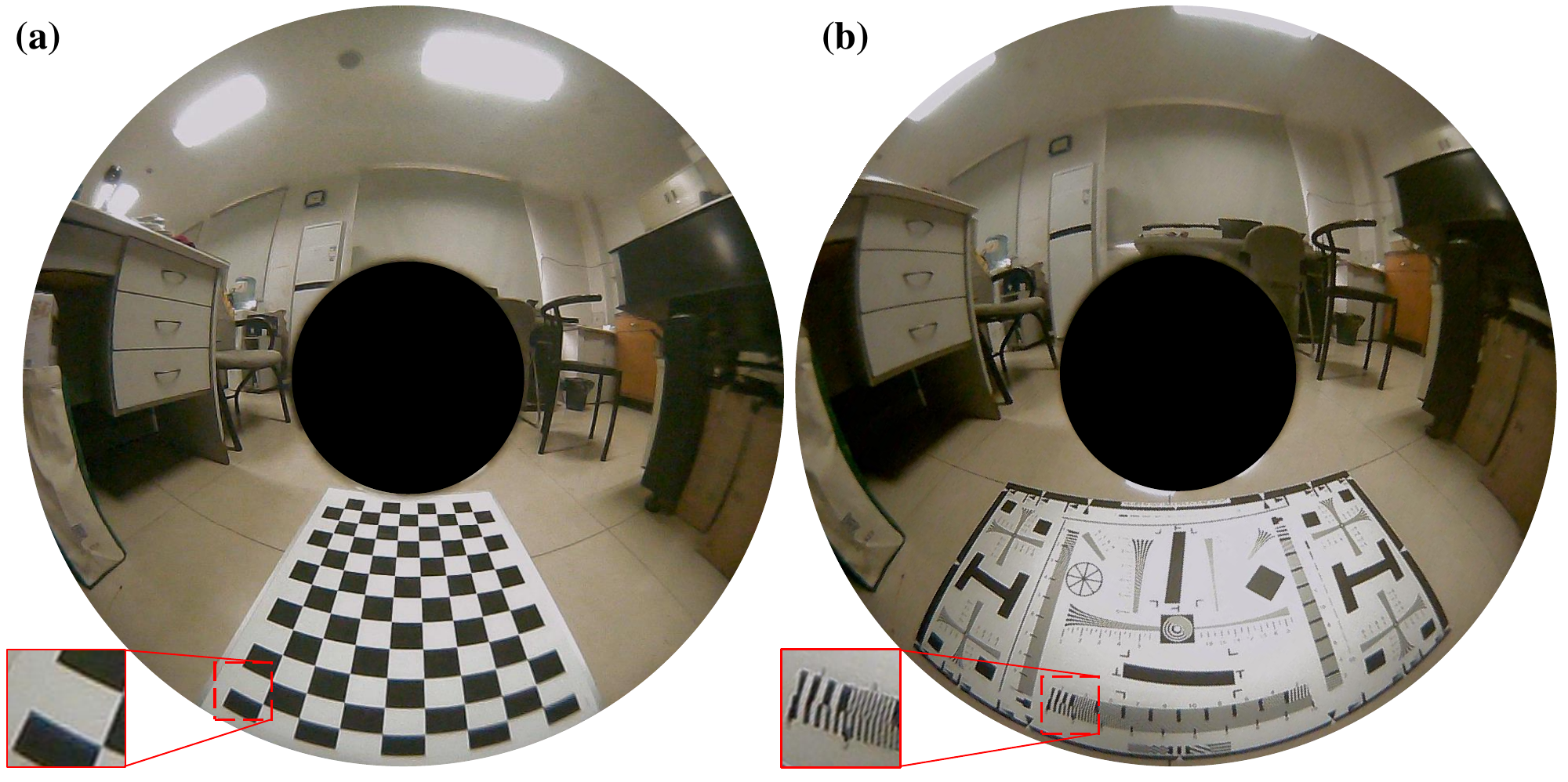}
\caption{Imaging performance test of the ASPAL. (a) Imaging result of the checkerboard pattern. (b) Imaging result of ISO12233 image resolution test card.}
\label{fig17}
\vskip-2ex
\end{figure}

For the manufactured ASPALs, we conduct imaging performance tests. As shown in Fig.~\ref{fig17}, a standard checkerboard pattern and an ISO12233 image resolution test card are taken at a height of 50 cm indoors using the ASPAL. The checkerboard pattern is a 12$\times$9 array with 5 cm sides, and the length and width of the ISO12233 image resolution test card are 75 cm$\times$43 cm. At 50 cm height, the ASPAL imaging result can resolve a 1.5 mm line width swept frequency wedge at the edge of the FoV, which demonstrates that the fabricated ASPAL has a good imaging performance. For the glass lenses in the ASPAL, we coated the antireflecting film on the transmitting surfaces and metal reflective film on the reflective surface. For plastic lenses, due to our low production quantities, the final lenses were not coated with antireflecting film. As a consequence, the rays from the ceiling light in Fig.~\ref{fig17} (a) are scattered. However, in theory, the plastic lenses should be coated with the antireflecting film to minimize the effects of stray light and increase transmittance.

\begin{figure}[ht!]
\centering\includegraphics[width=8.2cm]{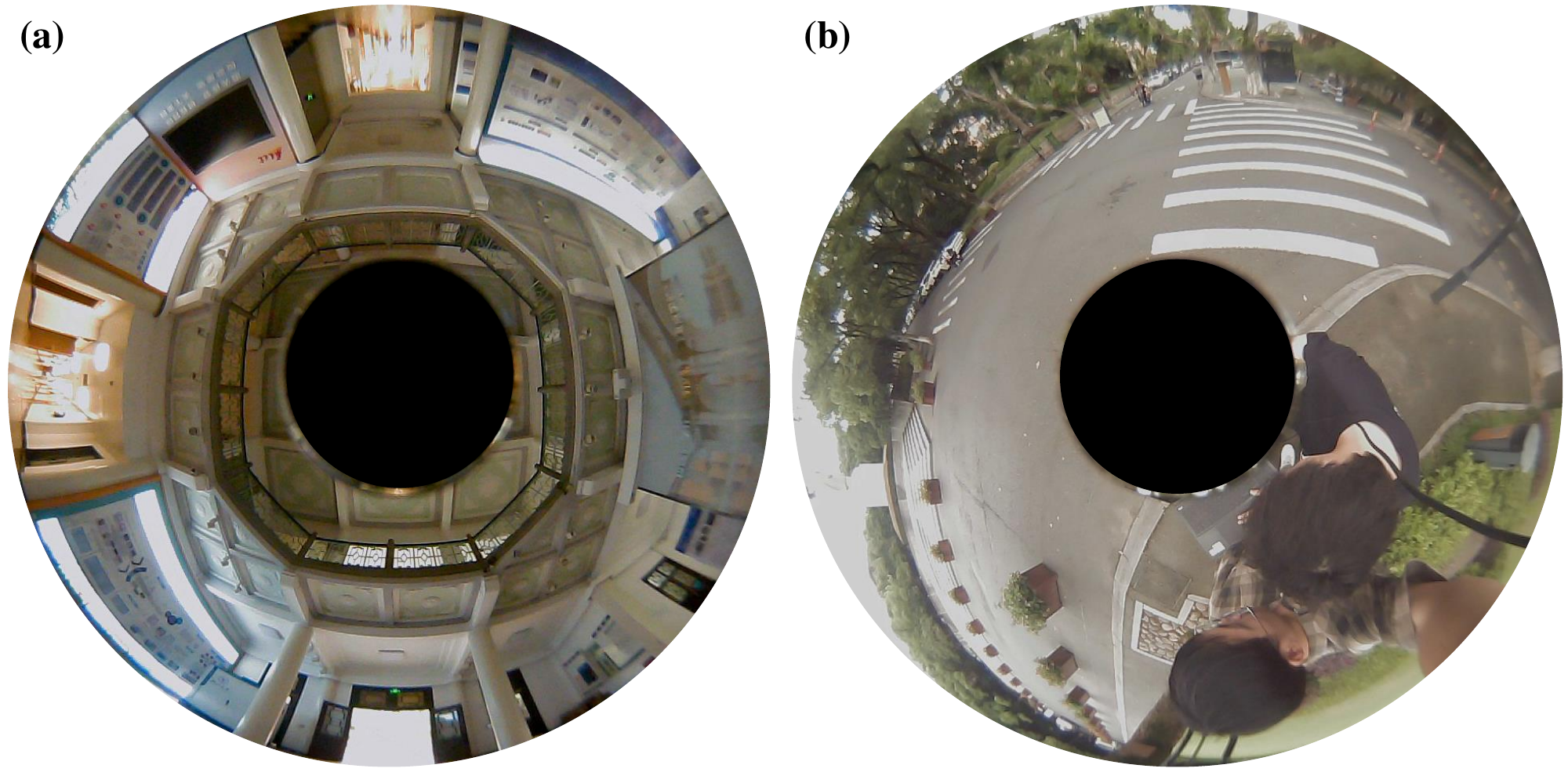}
\caption{Imaging performance test of the ASPAL. (a) The second-floor railing can be clearly distinguished in indoor scenes. (b) Traffic zebra crossing can be clearly distinguished in outdoor scenes.}
\label{fig18}
\vskip-3ex
\end{figure}

We also use the ASPAL for imaging testing of indoor and outdoor scenes. In Fig.~\ref{fig18}~(a), the railing on the second floor is distinguishable indoors. In Fig.~\ref{fig18}~(b), the ASPAL can obtain clear panoramic images with a large FoV in an outdoor traffic scene. The above experiment results demonstrate that the ASPAL proposed in this paper can be applied in the fields of miniaturized robot perception, miniaturized UAV perception, indoor and outdoor surveillance.

\section{Conclusion and future work}
The traditional panoramic imaging system has several major limitations such as large size, high weight, and complex system. To solve these drawbacks, we propose a high-performance glass-plastic hybrid minimalist ASPAL architecture design. Its focal length is -0.97 mm and the $F$$/$\# is 4.5. The FoV of the ASPAL is  360$^\circ$ $\times$(35$^\circ$ $\sim $110$^\circ$). The maximum spot RMS radius is less than 2.1 $\upmu$m, and the MTF reaches 0.47 at 133 lp/mm. In addition, we propose a tolerance analysis method applicable to ASPAL for the local tolerance requirements of the annular surfaces. This method constructs the Fringe Zernike model and the Standard Zernike model to simulate the irregularity of the aspheric surface, which can reasonably and accurately evaluate the PV and RMS figure error requirements of the optical working surface.

With the help of high-precision glass molding and injection molding aspheric lens manufacturing techniques, each manufactured ASPAL has only 8.5 g. Compared to traditional spherical PALs, our ASPAL has a larger panoramic FoV and advantages in volume and weight. Compared with traditional computational imaging methods used for minimalist optical systems, our proposed ASPAL system can achieve imaging quality close to the diffraction limit without requiring additional computing equipment and power consumption, reducing costs and system volume. Our ASPAL can obtain more realistic scene information than computational imaging. This study provides a low-cost and high-quality solution for the design, analysis, and fabrication of ultra-wide-angle panoramic systems that are minimalist and miniaturized. It enables the use of PALs in batch applications in space and weight-limited environmental sensing scenarios such as miniaturized drones and microrobots.
However, this study still has some potential for improvement. In the future, we aim to investigate new panoramic imaging systems that take into account compactness, high resolution, and large numerical aperture.
\vskip-2ex

\bibliographystyle{SELF-cas-model2-names}

\bibliography{ref}

\section*{Funding}

National Natural Science Foundation of China (NSFC) under Grant No. 12174341. Hangzhou SurImage Technology Company Ltd. Henan Province Key R\&D Special Project (2311111112700).

\end{document}